\documentclass[12pt,english,sort&compress, letterpaper]{JHEP3}
\usepackage[T1]{fontenc}
\usepackage{ae}
\usepackage{aecompl}
\usepackage[latin9]{inputenc}
\usepackage{amsthm}
\usepackage{amsmath}
\usepackage{amssymb}
\usepackage[numbers]{natbib}

\makeatletter
\theoremstyle{plain}

\title{Flux Attractors and Generating Functions}
\author{Finn Larsen and Ross O'Connell \\ 
Department of Physics, University of Michigan,\\
450 Church Street, Ann Arbor, MI 48109-1020, USA \\ 
Email: \email{larsenf@umich.edu}, \email{rcoconne@umich.edu}}
\abstract{We use the flux attractor equations to study IIB  supergravity compactifications with ISD fluxes. We show that the attractor equations determine not just the values of moduli fields, but also the masses of those moduli and the gravitino.  We then show that the flux attractor equations can be recast in terms of derivatives of a single generating function.  We also find a simple expression for this generating function in terms of the gravitino mass, with both quantities considered as functions of the fluxes. For a simple prepotential, we explicitly solve the attractor equations. We conclude by discussing a thermodynamic interpretation of this generating function, and possible implications for the landscape.}

\preprint{MCTP 09-15}

\makeatother

\usepackage{babel}

\begin{document}

\section{Introduction}

The compactification of string theory from 10 to 4 dimensions is a
subject of both formal and phenomenological interest. Many methods
of compactification result in moduli: massless 4D scalar fields which
correspond to deformations of the compactification geometry. Given
the observed absence of massless scalars, these moduli are phenomenologically
undesirable. As a result, much attention has been focused on the question
of how moduli can be \emph{stabilized}, i.e. how features can be added
to a simple compactification so that most or all of the 4D scalar
fields become massive. We can consider this question in three different
levels of detail:
\begin{enumerate}
\item Is the proposed stabilization method consistent? That is, does the
stabilized compactification still solve the 10D equations of motion?
\item Which moduli are stabilized, and what are their VEVs?
\item What are the masses of the moduli?
\end{enumerate}
In this paper we study compactifications of IIB string theory on Calabi-Yau
orientifolds, with RR and NS 3-form flux in the compact directions.
The \emph{flux attractor equations} \citep{Kallosh:2005ax} describing
the stabilization of the moduli strongly resemble black hole attractor
equations, and we will exploit this similarity to address the questions
above.

We will focus our attention on one of the 10D equations of motion.
If the (real) 3-form RR flux is $F_{3},$ the (real) 3-form NS flux
is $H_{3},$ and the complex axio-dilaton is $\tau,$ we define the
complex 3-form flux\begin{equation}
G_{3}\equiv F_{3}-\tau H_{3}\,.\label{eq:G3-def}\end{equation}
For large classes of compactifications to 4D Minkowski space, the
10D equations of motion require \citep{Dasgupta:1999ss,Giddings:2001yu} that $G_{3}$
be imaginary self dual (ISD):\begin{equation}
*_{6}G_{3}=iG_{3}\,.\label{eq:ISD}\end{equation}
Because $*_{6}$ involves the metric, a non-zero $G_{3}$ stabilizes
some or all of the complex structure moduli and $\tau.$ Specifically,
the complex structure of the Calabi-Yau is fixed so that $G_{3}$
has only $\left(0,3\right)$ and/or $\left(2,1\right)$ components.
If no such combination of complex structure and $\tau$ exists, the
choice of $F_{3}$ and $H_{3}$ is not consistent with compactification
to Minkowski space. In order to analyze \eqref{eq:ISD} in detail
we may expand $G_{3}$ and the holomorphic 3-form, $\Omega_{3},$
on a judiciously chosen basis of 3-cycles. This procedure results
in the flux attractor equations, as we review in section \ref{sec:From-ISD-Conditions}.

The resulting algebraic equations suffer an apparent inconsistency,
in that there are many more equations than moduli. If $n=b_{3}/2-1$
is the number of $\left(2,1\right)$ cycles on the Calabi-Yau, we
will find $4n+4$ different (real) equations and only $2n+2$ (real)
moduli. While this mismatch suggests that the system of equations
is overconstrained, we will show that this is not the case. In section
\ref{sec:Counting-equations-and} we will show that the $4n+4$ attractor
equations determine both the VEVs of the moduli \emph{and} the independent
parameters of their mass matrix, as well as the gravitino mass. All
of these outputs together constitute $4n+4$ parameters, the same
as the number of input fluxes.

Having established that the flux attractor equations determine both
the moduli VEVs\emph{ and} certain mass parameters, in section \ref{sec:Solving-the-Flux}
we develop an algorithm to find them. We take inspiration from OSV
\citep{Ooguri:2004zv}, who solved the black hole attractor equations
by introducing a mixed ensemble. Accordingly, we first solve the {}``magnetic''
half of the attractor equations, writing our $4n+4$ parameters in
terms of the $2n+2$ magnetic fluxes and $2n+2$ as-yet-undetermined
electric potentials. We then show that the {}``electric'' attractor
equations can be rewritten in terms of a generating function, and
that they can be formally solved by a simple Legendre transform. 

The existence of the generating function $\mathcal{G}$ is the principal
result of this paper. If one can determine it as a function of arbitrary
fluxes, its derivatives will give back the moduli VEVs and the mass
parameters. Thus $\mathcal{G}$ provides a compact summary of the
flux attractor behavior, and this suggests that we study the properties
of $\mathcal{G}$ directly. We initiate such a study in section \ref{sec:GeneralPropertiesOfG},
where we find a general formula for $\mathcal{G}:$\begin{equation}
\mathcal{G}=\int F_{3}\wedge H_{3}-2\mbox{Vol}^{2}m_{3/2}^{2}.\end{equation}
Here the gravitino mass is considered as a function of arbitrary fluxes. 

We proceed in section \ref{sec:STUSolution} by considering an explicit
example. We use the prepotential $F=Z^{1}Z^{2}Z^{3}/Z^{0},$ a setting
with sixteen distinct fluxes. For a reduced set of eight of these
fluxes we are able to completely solve the flux attractor equations.
We then argue that the general case can be solved as well, by appealing
to duality transformations.

For the sake of simplicity, we will discuss many of our results in
the context of large-volume, unwarped compactifications. These lead
to relatively well-understood 4D theories, and we can easily translate
our findings about the 10D geometry into statements about 4D physics.
However, our 10D reasoning applies equally well to strongly-warped
compactifications and some non-geometric compactifications \citep{Becker:2006ks}.
Since we are analyzing the ISD condition, which is quite robust, we
expect our qualitative understanding of the flux attractor behavior,
such as the existence of a generating function, to be similarly robust.
On the other hand the detailed mass spectrum depends on the Kähler
potential, and is therefore less robust.

As we have mentioned above, the solution of the flux attractor equations
is controlled by a single generating function, which depends on the
fluxes alone. In the case of the black hole attractor, the analogous
function turned out to the the equilibrium value of the black hole
entropy. It is tempting to speculate that the flux attractor equations
also describe a thermodynamic system. Ultimately, the underlying statistical
system may be related to a classical measure on this patch of the
string theory landscape. We conclude in section \ref{sec:Conclusion}
by summarizing the issues that must be resolved in order to make this
interpretation sound.

\section{\label{sec:From-ISD-Conditions}From the ISD Condition to Attractor
Equations}

In this section we review some basic aspects of special geometry and
flux compactifications. We then provide a simple derivation of the
flux attractor equations.

\subsection{\label{sub:Special-Geometry}Special Geometry}

Most of the objects we are interested in, including $F_{3},$ $H_{3},$
and $\Omega_{3},$ are 3-forms on the compact space. It is useful
to expand these 3-forms on a real basis $\left\{ \alpha_{I},\beta^{I}\right\} ,$
$I=0,...,n,$ satisfying\begin{eqnarray}
\int\alpha_{I}\wedge\beta^{J} & = & \delta_{I}^{J}\,,\label{eq:Ortho1}\\
\int\alpha_{I}\wedge\alpha_{J} & = & \int\beta^{I}\wedge\beta^{J}=0\,.\label{eq:Ortho2}\end{eqnarray}
We specify the NS fluxes $H_{3}$ and RR fluxes $F_{3}$ with respect
to this basis as\begin{eqnarray}
H_{3} & = & m_{h}^{I}\alpha_{I}-e_{I}^{h}\beta^{I}\,,\label{eq:Hab}\\
F_{3} & = & m_{f}^{I}\alpha_{I}-e_{I}^{f}\beta^{I}\,.\label{eq:Fab}\end{eqnarray}
 There is an $Sp\left(2n+2,\mathbb{R}\right)$ symmetry%
\footnote{Dirac quantization conditions require the magnetic fluxes $m_{h,f}^{I}$
and electric fluxes $e_{I}^{h,f}$ to take integer values, breaking
$Sp\left(2n+2,\mathbb{R}\right)$ to a discrete subgroup.%
} that corresponds to a change in the basis $\left\{ \alpha_{I},\beta^{I}\right\} .$
The fluxes $\left\{ m_{h}^{I},e_{I}^{h}\right\} $ and $\left\{ m_{f}^{I},e_{I}^{f}\right\} $
transform in the fundamental of $Sp\left(2n+2,\mathbb{R}\right),$
and objects with an index $I,J,K...$ transform in the fundamental
of $SO\left(n+1,\mathbb{R}\right)\subset Sp\left(2n+2,\mathbb{R}\right).$ 

We can also expand the holomorphic 3-form with respect to the real
basis,\begin{equation}
\Omega_{3}=Z^{I}\alpha_{I}-F_{I}\beta^{I}\,.\label{eq:Oab}\end{equation}
The combination $\left\{ Z^{I},F_{I}\right\} $\textbf{ }is called
a \emph{symplectic section }\citep{Ceresole:1995ca}, and also transforms
in the fundamental of $Sp\left(2n+2,\mathbb{R}\right)$. While the
fluxes $e_{I}^{h,f}$ and $m_{h,f}^{I}$ were all independent parameters,
the $F_{I}$ and $Z^{I}$ are holomorphic functions of the complex
structure moduli. For our purposes, it is sufficient to treat the
$F_{I}$ as functions that are holomorphic and homogeneous of degree
1 in the $Z^{I}.$ The functional form of the $F_{I}$ is the only
information about the Calabi-Yau geometry that we will use. 

The holomorphic 3-form is only defined up to a holomorphic rescaling,\begin{equation}
\Omega_{3}\to f\left(Z^{I}\right)\Omega_{3}\,.\end{equation}
These are the \emph{Kähler transformations. }If, under Kähler transformations,
an operator is simply multiplied by $h$ powers of $f\left(Z^{I}\right)$
and $\overline{h}$ powers of $\overline{f\left(Z^{I}\right)},$ we
will say that it is Kähler covariant with weight $\left(h,\overline{h}\right).$
For example, $\Omega_{3}$ has weight $\left(1,0\right).$ 

Physical moduli must be invariant under Kähler transformations. For
example, on a patch where $Z^{0}\neq0$ we may use the ratios\begin{equation}
z^{i}\equiv\frac{Z^{i}}{Z^{0}}\,,\end{equation}
where $i=1,...,n.$ The $z^{i}$ are clearly Kähler invariant. Unfortunately,
this breaks the $SO\left(n+1\right)$ symmetry enjoyed by the $Z^{I},$
so we will sometimes use an alternative approach to formulating Kähler
invariant quantities. We will utilize a coefficient $C$ which has
weight $\left(-1,0\right),$ so that the products $CZ^{I}$ are Kähler
invariant.

Because Kähler transformations are local, ordinary derivatives of
Kähler covariant functions do not give new Kähler covariant functions.
We introduce the Kähler potential\begin{equation}
K_{z}=-\log i\int\Omega_{3}\wedge\overline{\Omega}_{3}\,,\label{eq:K-z}\end{equation}
which generates the metric on moduli space,\begin{eqnarray}
g_{i\overline{j}} & = & \partial_{i}\overline{\partial}_{\overline{j}}K_{z}\,.\label{eq:KMetric}\end{eqnarray}
By construction, $e^{K_{z}}$ has weight $\left(-1,-1\right).$ This
motivates the definition of the Kähler covariant derivative of an
operator of weight $\left(h,\overline{h}\right),$\begin{equation}
D_{i}\mathcal{O}^{\left(h,\overline{h}\right)}\equiv e^{-hK_{z}}\partial_{i}\left(e^{hK_{z}}\mathcal{O}^{\left(h\overline{h}\right)}\right)=\partial_{i}\mathcal{O}\left(h,\overline{h}\right)+h\mathcal{O}^{\left(h,\overline{h}\right)}\partial_{i}K_{z}\,.\end{equation}
We note that since the Kähler potential is real, the Kähler covariant
derivative of a holomorphic object is not itself holomorphic. 

It is especially interesting to consider derivatives of the holomorphic
3-form. An ordinary derivative with respect to the complex structure
moduli gives a sum of $\left(3,0\right)$ and $\left(2,1\right)$
forms,\begin{equation}
\partial_{i}\Omega_{3}=k_{i}\Omega_{3}+\chi_{i}\,.\label{eq:dOmega}\end{equation}
If we instead use a Kähler covariant derivative, the Kähler potential
is constructed so that the $\left(3,0\right)$ piece cancels and we
are left with only a $\left(2,1\right)$ form, \begin{equation}
D_{i}\Omega_{3}=\chi_{i}\,.\label{eq:Omega-Derivs}\end{equation}
This establishes a convenient \emph{complex} basis for 3-forms on
the Calabi-Yau, $\{\Omega_{3},$ $D_{i}\Omega_{3},$ $\overline{D_{i}\Omega}_{3},$
$\overline{\Omega}_{3}\}$ \citep{Candelas:1990pi}. The intimate
connection between the complex structure of a Calabi-Yau and its cohomology
will be the primary tool that we use to analyze the ISD condition
\eqref{eq:ISD}.

\subsection{S-Duality}

In addition to Kähler transformations, S-duality helps organize the
flux attractor equations. Type IIB supergravity has an $SL\left(2,\mathbb{R}\right)$
symmetry%
\footnote{Quantum effects break this to $SL\left(2,\mathbb{Z}\right),$ but
the distinction between the two groups will not be relevant to our
analysis.%
}, under which

\begin{eqnarray}
\tau & \to & \frac{a\tau+b}{c\tau+d}\,,\\
\left(\begin{array}{c}
F_{3}\\
H_{3}\end{array}\right) & \to & \left(\begin{array}{cc}
a & b\\
c & d\end{array}\right)\left(\begin{array}{c}
F_{3}\\
H_{3}\end{array}\right),\end{eqnarray}
with the constraint \begin{equation}
ad-bc=1\,.\end{equation}
The transformation of the complex flux $G_{3}$ under S-duality can
be deduced from the transformations of $F_{3},$ $H_{3},$ and $\tau:$\begin{equation}
G_{3}\to\frac{G_{3}}{c\tau+d}\,.\end{equation}
We will frequently encounter $\mbox{Im}\left(\tau\right),$ which
transforms as\begin{equation}
\mbox{Im}\left(\tau\right)\to\frac{\mbox{Im}\left(\tau\right)}{\left|c\tau+d\right|^{2}}\,.\label{eq:Imtau-S-tform}\end{equation}

\subsection{\label{sub:Large-Volume}4D Physics of Large Volume Compactifications}

The flux attractor equations are simply a rephrasing of the ISD condition
\eqref{eq:ISD}. We could discuss the ISD condition entirely from
the 10D point of view, but we find it useful to make reference to
the resulting 4D effective theory. As long as the volume of the Calabi-Yau
is large relative to the string scale, and regions of strong warping
are all string scale, the result is a 4D, $\mathcal{N}=1$ theory
with the GVW superpotential \citep{Gukov:1999ya,Taylor:1999ii}, \begin{equation}
W=\int_{CY}G_{3}\wedge\Omega_{3}\,,\label{eq:GVW}\end{equation}
and Kähler potential%
\footnote{The volume of the Calabi-Yau is determined by the Kähler moduli, which
are not stabilized by 3-form fluxes. We have little to say about the
factors of the volume that appear, but include them for completeness. %
}, \begin{eqnarray}
K & = & K_{z}+K_{\tau}+K_{t}\\
 & = & -\log\left[i\int_{CY}\Omega_{3}\wedge\overline{\Omega}_{3}\right]-\log\left[2\mbox{Im}\left(\tau\right)\right]-2\log\left[\mbox{Vol}\right]\,.\label{eq:KahlerPot}\end{eqnarray}
These compactifications are reviewed in e.g. \citep{Silverstein:2004id,Grana:2005jc,Douglas:2006es,Denef:2008wq}.
While both the superpotential and Kähler potential receive a variety
of phenomenologically interesting corrections \citep{Kachru:2003aw,Balasubramanian:2005zx,Berg:2007wt},
we will not consider their effects here. Note that the 4D Kähler potential
contains the Kähler potential \eqref{eq:K-z} that we introduced earlier
for the Calabi-Yau. This relationship between the 4D kinetic terms
and the Calabi-Yau geometry is a special characteristic of the large-volume
limit, and breaks down in the presence of significant warping (see e.g. \citep{DeWolfe:2002nn,Frey:2002hf,Giddings:2005ff,Frey:2006wv,Shiu:2008ry,Frey:2008xw}).

In addition to the complex structure moduli $z^{i}$ and $\tau,$
the 4D theory also contains a number of Kähler moduli $t^{a}.$ Rather
than depending on the holomorphic volumes of three-cycles, these measure
the actual volumes of two- and four-cycles. Since the Kähler moduli
do not appear in the superpotential, their F-terms are just\begin{equation}
F_{a}=D_{a}W=W\partial_{a}K_{t}\,.\label{eq:Kahler-F-Terms}\end{equation}
 When summed up they give $\sum_{a}\left|F_{a}\right|^{2}=3\left|W\right|^{2},$
so the standard expression for the scalar potential simplifies to

\begin{eqnarray}
V & = & e^{K}\left[\sum_{A=i,\tau,a}\left|D_{A}W\right|^{2}-3\left|W\right|^{2}\right]\label{eq:NoScalePotential}\\
 & = & e^{K}\left(\sum_{i}\left|D_{i}W\right|^{2}+\left|D_{\tau}W\right|^{2}\right).\end{eqnarray}
When $W\neq0$ the F-terms for the Kähler moduli \eqref{eq:Kahler-F-Terms}
are non-vanishing, so SUSY is broken. However, the potential \eqref{eq:NoScalePotential}
is positive definite and has a global minimum when $F_{i}=D_{i}W=0$
and $F_{\tau}=D_{\tau}W=0.$ Because of this, we require that $F_{i}=F_{\tau}=0,$
regardless of whether SUSY is broken. 

The simple form of the Kähler potential gives the $F_{i}=F_{\tau}=0$
conditions simple geometric interpretations. For the complex structure
moduli we find\begin{eqnarray}
D_{i}W & = & \int_{CY}G_{3}\wedge D_{i}\Omega_{3}=\int_{CY}G_{3}\wedge\chi_{i}\,,\label{eq:Di}\end{eqnarray}
 so that setting $F_{i}=0$ is equivalent to requiring that $G_{3}$
have no $\left(1,2\right)$ component. In addition one can verify
that\begin{equation}
D_{\tau}\int G_{3}\wedge\Omega_{3}=-\frac{1}{\tau-\overline{\tau}}\int\overline{G}_{3}\wedge\Omega_{3}\,,\label{eq:Dtau}\end{equation}
so setting $F_{\tau}=0$ is equivalent to requiring that $G_{3}$
have no $\left(3,0\right)$ component. Thus we have found that minimizing
the potential \eqref{eq:NoScalePotential} is \emph{equivalent} to
imposing the ISD condition \eqref{eq:ISD}. This is one of the reasons
that the GVW superpotential is believed to accurately describe large-volume
compactifications.

\subsection{\label{sub:Flux-Attractor-Equations}Flux Attractor Equations}

The flux attractor equations were originally derived in \citep{Kallosh:2005ax}
by considering F-theory compactified on $\mbox{CY}_{3}\times T^{2}.$
For the sake of variety, we present a slightly different derivation
which does not involve an explicit embedding in F-theory.

Our goal is to make the implications of the ISD condition \eqref{eq:ISD}
more explicit. Since an ISD 3-form can have only $\left(0,3\right)$
and $\left(2,1\right)$ pieces, we can expand it with respect to the
complex basis introduced at the end of section \eqref{sub:Special-Geometry}
as:\begin{equation}
G_{3}=-i\mbox{Im}\left(\tau\right)\left[\overline{C}\overline{\Omega}_{3}+C^{i}D_{i}\Omega_{3}\right]\,.\label{eq:HelloC^i}\end{equation}
The overall factor of $-i\mbox{Im}\left(\tau\right)$ is included
for convenience. Note that $C$ and $C^{i}$ both have weight $\left(-1,0\right)$
under Kähler transformations, and transform under S-duality as\begin{eqnarray}
C & \to & \left(c\tau+d\right)C\,,\label{eq:Coeffs-S-tform}\\
C^{i} & \to & \left(c\overline{\tau}+d\right)C^{i}\,.\label{eq:Coeffs-S-tform-2}\end{eqnarray}
In order to make \eqref{eq:HelloC^i} completely explicit we must
specify the symplectic section $\left\{ Z^{I},F_{I}\right\} ,$ as
this determines how $\Omega_{3}$ depends on the complex structure
moduli. We can then compute the Kähler covariant derivatives $D_{i}\Omega_{3},$
so that \eqref{eq:HelloC^i} becomes an \emph{algebraic} equation
for the complex structure moduli and the axio-dilaton.

One undesirable aspect of \eqref{eq:HelloC^i} is that the LHS contains
both the real fluxes $F_{3}$ and $H_{3},$ which we think of us {}``inputs,''
and the axio-dilaton $\tau,$ which we think of as an {}``output.''
This is rectified by writing\begin{equation}
\left(\begin{array}{c}
G_{3}\\
\overline{G}_{3}\end{array}\right)=\left(\begin{array}{cc}
1 & -\tau\\
1 & -\overline{\tau}\end{array}\right)\left(\begin{array}{c}
F_{3}\\
H_{3}\end{array}\right)=-i\mbox{Im}\left(\tau\right)\left(\begin{array}{c}
\overline{C}\overline{\Omega}_{3}+C^{i}D_{i}\Omega_{3}\\
-C\Omega_{3}-\overline{C}^{i}\overline{D_{i}\Omega}_{3}\end{array}\right),\end{equation}
which we can easily invert:\begin{eqnarray}
\left(\begin{array}{c}
F_{3}\\
H_{3}\end{array}\right) & = & -\frac{1}{2}\left(\begin{array}{cc}
-\overline{\tau} & \tau\\
-1 & 1\end{array}\right)\left(\begin{array}{c}
\overline{C}\overline{\Omega}_{3}+C^{i}D_{i}\Omega_{3}\\
-C\Omega_{3}-\overline{C}^{i}\overline{D_{i}\Omega}_{3}\end{array}\right)\\
 & = & \left(\begin{array}{c}
\mbox{Re}\left[\tau\left(C\Omega_{3}+\overline{C}^{i}\overline{D_{i}\Omega}_{3}\right)\right]\\
\mbox{Re}\left[C\Omega_{3}+\overline{C}^{i}\overline{D_{i}\Omega}_{3}\right]\end{array}\right).\label{eq:FormEqs}\end{eqnarray}
Now the LHS of the attractor equations consists entirely of quantities
that define the vacuum (fluxes), while the RHS depends on the moduli
and the symplectic section (choice of $\left\{ Z^{I},F_{I}\right\} $).

The equations in \eqref{eq:FormEqs} are equations for 3-forms, rather
than for ordinary numbers. While this makes their geometric implications
clear, if we want to actually solve the equations it will be helpful
to integrate them against a real basis of 3-forms. We have already
introduced the required notation in \eqref{eq:Hab}-\eqref{eq:Oab},
so we simply quote the result,\begin{eqnarray}
m_{f}^{I} & = & \mbox{Re}\left[\tau\left(CZ^{I}+\overline{C}^{i}\overline{D_{i}Z^{I}}\right)\right],\label{eq:FullAttEqsC1}\\
m_{h}^{I} & = & \mbox{Re}\left[CZ^{I}+\overline{C}^{i}\overline{D_{i}Z^{I}}\right],\label{eq:FullAttEqsC2}\\
e_{I}^{f} & = & \mbox{Re}\left[\tau\left(CF_{I}+\overline{C}^{i}\overline{D_{i}F_{I}}\right)\right],\label{eq:FullAttEqsC3}\\
e_{I}^{h} & = & \mbox{Re}\left[CF_{I}+\overline{C}^{i}\overline{D_{i}F_{I}}\right].\label{eq:FullAttEqsC4}\end{eqnarray}
One benefit to writing the attractor equations in this form is that
there is manifestly one real equation for each real flux, for a total
of $4n+4$ real equations. We will compare this to the number of moduli
and other parameters quite carefully in the next section.

One may wonder to what extent it makes sense to call \eqref{eq:FullAttEqsC1}-\eqref{eq:FullAttEqsC4}
{}``attractor equations.'' The word {}``attractor'' implies some
sort of flow along which all information about a set of initial conditions
is lost, but we have not introduced any notion of attractor flow.
We note that in the study of extremal black holes, there is a useful
distinction between the entire attractor \emph{flow},\emph{ }which
takes place between spatial infinity and the horizon,\emph{ }and the
attractor \emph{equations}, which describe how the moduli are stabilized
at the horizon. Because \eqref{eq:FullAttEqsC1}-\eqref{eq:FullAttEqsC4}
are closely analogous to the black hole attractor equations, we consider
calling them {}``attractor equations'' to be only a minor abuse
of the term.

\section{\label{sec:Counting-equations-and}Attractor Equations and Mass Matrices}

In expanding out the flux attractor equations, we found $4n+4$ real
equations%
\footnote{$n=b_{3}/2-1,$ so that $n+1$ is the number of $\mathcal{N}=1$ vector
multiplets in the 4D theory.%
} \eqref{eq:FullAttEqsC1}-\eqref{eq:FullAttEqsC4}. This is many more
than the $2n+2$ real moduli VEVs we want to fix, the $z^{i}$ and
$\tau.$ The origin of this mismatch is that there are additional
{}``outputs'' of the attractor equations, namely the coefficients
$C$ and $C^{i}.$ Including these outputs gives $4n+4$ real variables,
equal to the number of attractor equations. We will see that these
coefficients determine the mass spectrum of the 4D theory.

\subsection{\label{sub:Black-Hole}Black Hole Attractor Equations and the Entropy}

While the $C^{i}$ are a new feature of the flux attractor equations,
the coefficient $C$ also appears in the more familiar context of
BPS black hole attractor equations. We begin by discussing the role
it plays there. Suppose we have constructed a 4D BPS Reissner-Nordström
black hole by wrapping D3 branes on the 3-cycles of a Calabi-Yau manifold.
The charges of the black hole can be described by a 3-form, $F_{3}.$
We can expand a general real 3-form either against a real basis, or
against the complex basis introduced in section \ref{sub:Special-Geometry}:\begin{eqnarray}
F_{3} & = & p^{I}\alpha_{I}-q_{I}\beta^{I}\\
 & = & \mbox{Re}\left[C\Omega_{3}+C^{i}D_{i}\Omega_{3}\right]\,.\label{eq:Generic-F}\end{eqnarray}
The expression for the spacetime central charge of the black hole
is\begin{equation}
W_{BH}=\int F_{3}\wedge\Omega_{3}\,,\label{eq:BHCentralCharge}\end{equation}
and the BPS conditions are $D_{i}W_{BH}=0.$ Since $F_{3}$ does not
depend on the moduli, the BPS conditions reduce to\begin{equation}
\int F_{3}\wedge D_{i}\Omega_{3}=0\,,\end{equation}
i.e. they require that the $\left(1,2\right)$ piece of $F$ vanishes.
This simplifies the general expansion \eqref{eq:Generic-F} to \begin{equation}
F_{3}=2\mbox{Re}\left[C\Omega_{3}\right]\,.\label{eq:BHAtt}\end{equation}
This is the standard black hole attractor equation, originally derived
in \citep{Ferrara:1995ih,Strominger:1996kf,Ferrara:1996um} and reviewed
in \citep{Moore:1998pn,Sen:2007qy,Bellucci:2007ds}.

If we expand \eqref{eq:BHAtt} on the real basis $\left\{ \alpha_{I},\beta^{I}\right\} $
we will find a counting problem. Although there are $2n+2$ real equations,
there are only $2n$ real physical moduli, the $z^{i}.$ In order
to understand the mismatch, we first note that the righthand side
of \eqref{eq:BHAtt} contains $2n+4$ real parameters, $\left\{ C,Z^{I}\right\} .$
Since both $C$ and $Z^{I}$ transform under Kähler transformations
we can eliminate one complex parameter, leaving $2n+2$ Kähler invariant
parameters. For example, if we assume that $Z^{0}\neq0,$ we can take
the Kähler invariant parameters to be $\left\{ CZ^{0},z^{i}=Z^{i}/Z^{0}\right\} .$
More generally, the number of Kähler invariant parameters is equal
to the number of attractor equations. The non-trivial feature is that,
in addition to determining the values of the moduli $z^{i},$ the
black hole attractor equations fix the Kähler invariant quantity $CZ^{0}.$ 

It is natural to ask what the physical significance of the additional
parameter is. One important place where it appears is in the black
hole entropy,\begin{eqnarray}
\frac{S}{\pi} & = & e^{K_{z}}\left|W_{BH}\right|^{2}\\
 & = & \frac{e^{-K_{z}}}{\left|Z^{0}\right|^{2}}\cdot\left|CZ^{0}\right|^{2}\,,\end{eqnarray}
since \eqref{eq:BHCentralCharge} and \eqref{eq:BHAtt} imply that
$W_{BH}=-i\overline{C}e^{-K_{z}}.$ In the final expression we have
written the black hole entropy as the product of two Kähler-invariant
factors, with the first factor depending only on the moduli $z^{i}.$
We see that a change in $CZ^{0}$ leads to a change in the entropy,
with the moduli held fixed. 

It is sometimes stated that solving the attractor equations is equivalent
to minimizing an effective potential. Our analysis shows that, in
fact, the attractor equations simultaneously determine both the values
of the moduli \emph{and} the value of the effective potential. Simply
minimizing the effective potential with respect to the moduli would
have given us $2n$ real equations, rather than $2n+2,$ and we would
have had to insert the solutions for the moduli back into the effective
potential to find its value at the minimum.

\subsection{Fermion Masses}

Let us now return to the flux attractor equations. \eqref{eq:FullAttEqsC1}-\eqref{eq:FullAttEqsC4}
constitute $4n+4$ real equations, while the moduli $z^{i}$ and $\tau$
constitute $2n+2$ real parameters. Our analysis of the black hole
attractor equations revealed that $CZ^{0}$ contributes two more real
independent parameters, but we are still left with $2n$ more equations
than parameters. The new features in the flux attractor are the coefficients
$C^{i},$ first introduced in \eqref{eq:HelloC^i}. Including these
in our set of Kähler-invariant parameters as $\left\{ \tau,z^{i},CZ^{0},C^{i}Z^{0}\right\} ,$
we have accounted for everything that appears on the righthand side
of \eqref{eq:FormEqs}, for a grand total of $4n+4$ parameters. Just
as in the black hole case we found that different choices of charges
could lead to the same moduli but different entropies, here different
choices of the fluxes can lead to the same moduli, but different values
of $CZ^{0}$ \emph{and} $C^{i}Z^{0}.$

In large-volume compactifications, the role of the black hole entropy
is played by the gravitino mass:\begin{equation}
m_{3/2}^{2}=e^{K}\left|W\right|^{2},\label{eq:GraivitinoDef}\end{equation}
Indeed, if we substitute in the expressions \eqref{eq:GVW} for the
superpotential and \eqref{eq:KahlerPot} for the Kähler potential,
we find \begin{equation}
m_{3/2}^{2}=\frac{e^{-K_{z}}\mbox{Im}\left(\tau\right)}{2\left|Z^{0}\right|^{2}\mbox{Vol}^{2}}\cdot\left|CZ^{0}\right|^{2}.\label{eq:FaveGravitino}\end{equation}
Just as $CZ^{0}$ determined the entropy of the black hole attractor,
it determines the gravitino mass for the flux attractor.

While we understand well enough what it means to solve for the VEVs
of $z^{i}$ and $\tau,$ and we know that $C$ is related to the gravitino
mass, we need to develop a physical interpretation of the $C^{i}.$
We'll first observe that the $C^{i}$ appear when we consider the
second derivatives of the superpotential:\begin{eqnarray}
D_{i}D_{j}W & = & \int G_{3}\wedge D_{i}D_{j}\Omega_{3}\\
 & = & \int G_{3}\wedge\left(\mathcal{F}_{ijk}\overline{\chi}^{k}\right)\\
 & = & \mbox{Im\ensuremath{\left(\tau\right)}}e^{-K_{z}}\mathcal{F}_{ijk}C^{k}\,,\label{eq:Dij}\end{eqnarray}
where \citep{Denef:2004ze}\textbf{ \begin{equation}
\mathcal{F}_{ijk}=ie^{K_{z}}\int\Omega_{3}\wedge\partial_{i}\partial_{j}\partial_{k}\Omega_{3}\end{equation}
}depends on both the moduli and the symplectic section%
\footnote{For cubic prepotentials and physical moduli $z^{i}=Z^{i}/Z^{0},$
$\int\Omega_{3}\wedge\partial_{i}\partial_{j}\partial_{k}\Omega_{3}=\left(Z^{0}\right)^{2}C_{ijk}$.%
}. We also need the mixed derivatives,\begin{eqnarray}
D_{\tau}D_{i}W & = & -\frac{\int\overline{G}_{3}\wedge\chi_{i}}{\tau-\overline{\tau}}\\
 & = & -\frac{1}{2}\int\left(C\Omega_{3}+\overline{C^{j}\chi_{j}}\right)\wedge\chi_{i}\\
 & = & \frac{i}{2}\overline{C}^{\overline{j}}g_{i\overline{j}}e^{-K_{z}}\,.\label{eq:Ditau}\end{eqnarray}
Here we used \eqref{eq:Di} and \eqref{eq:Dtau}. Also, in the last
step we used the relationship between the metric on complex structure
moduli space \eqref{eq:KMetric} and the $\left(2,1\right)$ forms
\eqref{eq:dOmega},\begin{equation}
g_{i\overline{j}}=-\frac{\int\chi_{i}\wedge\overline{\chi}_{\overline{j}}}{\int\Omega_{3}\wedge\overline{\Omega}_{3}}\,.\label{eq:21metric}\end{equation}
The remaining second derivative vanishes,\begin{eqnarray}
D_{\tau}D_{\tau}W & = & \frac{2}{\left(\tau-\overline{\tau}\right)^{2}}\int\overline{G}_{3}\wedge\Omega_{3}\\
 & = & 0\,,\label{eq:Dtautau}\end{eqnarray}
since $\overline{G}_{3}$ has no $\left(0,3\right)$ piece. 

The second derivatives of the superpotential generically determine
the masses of the components of chiral multiplets. The standard expression
\citep{Wess:1992cp} for the spinor mass matrix in 4D $\mathcal{N}=1$
supergravity is\begin{equation}
m_{\alpha\beta}=\left(D_{\alpha}D_{\beta}W-\frac{2}{3}\left(D_{\alpha}W\right)\left(D_{\beta}W\right)-\Gamma_{\alpha\beta}^{c}D_{c}W\right)\frac{m_{3/2}}{W}\,.\end{equation}
Since the Kähler moduli are not stabilized, we will only consider
$\alpha=i,\tau.$ The moduli space factorizes, so the connection $\Gamma_{BC}^{A}$
will have no mixed components, $\Gamma_{\alpha\beta}^{a}=0.$ Imposing
the global minimum condition $D_{i}W=D_{\tau}W=0$ reduces the mass
matrix to\begin{equation}
m_{\alpha\beta}=e^{K/2}\sqrt{\frac{\overline{W}}{W}}D_{\alpha}D_{\beta}W\,.\label{eq:Fermion}\end{equation}
Note that the overall phase $\sqrt{\overline{W}/W}$ could be absorbed
into the definition of the fermions, though we will not do so here.
Substituting in the second derivatives computed above, the fermion
mass matrix simplifies to\begin{equation}
\left(\begin{array}{cc}
m_{ij} & m_{i\tau}\\
m_{\tau i} & m_{\tau\tau}\end{array}\right)=\frac{m_{3/2}}{\overline{C}}\left(\begin{array}{cc}
\mathcal{F}_{ijk}C^{k} & -\frac{1}{2i\mathrm{Im}\left(\tau\right)}g_{i\overline{j}}\overline{C}^{\overline{j}}\\
-\frac{1}{2i\mathrm{Im}\left(\tau\right)}g_{i\overline{j}}\overline{C}^{\overline{j}} & 0\end{array}\right).\label{eq:MassMatrix}\end{equation}
Here we used \eqref{eq:Fermion} and \eqref{eq:GraivitinoDef}, substituted
in the second derivatives \eqref{eq:Dij}, \eqref{eq:Ditau}, and
\eqref{eq:Dtautau}, then simplified using \eqref{eq:K-z}, \eqref{eq:GVW},
\eqref{eq:HelloC^i}, and \eqref{eq:21metric}. This demonstrates
how, in the large volume scenario, the $C^{i}$ determine the structure
of the fermion mass matrix. These masses remain finite even in the
limit $m_{3/2}\sim\left|C\right|\to0,$ since the ratio $m_{3/2}/\overline{C}$
approaches a finite value.

A few comments are in order. First, the fermion mass matrix has $2n+2$
real eigenvalues, two more than there are parameters $C^{i}.$ This
indicates that we cannot independently determine the masses of \emph{all}
of the moduli -- for example, we could consider choosing the masses
of the $z^{i},$ but then the mass of $\tau$ would be determined.
It is also interesting that the form of $m_{ij}$ suggests a generalized
Higgs mechanism. If we think of the $\mathcal{F}_{ijk}$ as Yukawa
couplings, than $C^{k}$ appears to play the role of a Higgs vacuum
expectation value. While the $C^{k}$ do not correspond to the expectation
values of any dynamical scalars, it is possible that they can be interpreted
as the expectation values of auxiliary fields. Finally, if we can
make $\mbox{Im}\left(\tau\right)=1/g_{s}$ large, then the smallest
fermion mass will be roughly $m_{3/2}g_{s}^{2}.$ It would be interesting
to see if such a light mode is of phenomenological interest, perhaps
at an intermediate scale.

\subsection{Scalar Masses}

In supersymmetric vacua, the masses of scalar fields should match
the masses of their fermionic partners. However, the no-scale vacua
that we consider generically break supersymmetry. While the F-terms
for the complex structure moduli and axio-dilaton vanish, $D_{i}W=D_{\tau}W=0$,
the F-terms for the Kähler moduli only vanish when $W=0,$ as shown
in \eqref{eq:Kahler-F-Terms}. In this case, the scalar mass-squared
matrix takes the following form:\begin{eqnarray}
\mathcal{M}^{2} & = & \left(\begin{array}{cc}
M_{\alpha\beta} & M_{\alpha\overline{\beta}}\\
\overline{M}_{\overline{\alpha}\beta} & \overline{M}_{\overline{\alpha\beta}}\end{array}\right),\\
M_{\alpha\beta}^{2} & = & e^{K}\overline{W}\left(D_{\alpha}D_{\beta}W+D_{\beta}D_{\alpha}W\right),\label{eq:Scalar1}\\
M_{\alpha\overline{\beta}}^{2} & = & e^{K}\left[g^{\gamma\overline{\delta}}D_{\alpha}D_{\gamma}W\overline{D_{\beta}D_{\delta}W}+\left|W\right|^{2}g_{\alpha\overline{\beta}}\right].\label{eq:Scalar2}\end{eqnarray}
While \eqref{eq:Scalar1} and \eqref{eq:Scalar2} would be standard
expressions for a theory with \emph{only} the complex structure moduli
and axio-dilaton, we verify in appendix \ref{sec:Scalar-Mass-Matrix}
that they also hold when Kähler moduli are included, and supersymmetry
is broken in that sector. Note that when $W=0,$ i.e. when supersymmetry
is preserved, $M_{\alpha\beta}^{2}$ vanishes and $M_{\alpha\overline{\beta}}^{2}=g^{\gamma\overline{\delta}}m_{\alpha\gamma}\overline{m}_{\overline{\delta\beta}},$
as expected. When $W\neq0,$ the scalar masses are lifted above the
fermion masses, and the splitting of the masses-squared is of order
$m_{3/2}^{2}=e^{K}\left|W\right|^{2}\sim\left|CZ^{0}\right|^{2}.$

\section{\label{sec:Solving-the-Flux}A Generating Function for the Flux Attractor
Equations}

In this section we develop an algorithm which, in principle, solves
the flux attractor equations. To do so we adapt the OSV solution of
the black hole attractor equations \citep{Ooguri:2004zv}. We begin
with a change of variables designed to automatically solve the magnetic
half of the attractor equations. Next, we rewrite the electric half
of the attractor equations as derivatives of a generating function.
Finally, a Legendre transform provides a formal solution of the attractor
equations. 

The generating function itself is quite interesting. In \citep{Ooguri:2004zv},
the generating function governing the black hole attractor turned
out to be the free energy of the black hole. Our interest in the generating
function is not restricted to this section, rather we will discuss
some of its general properties in section \ref{sec:GeneralPropertiesOfG}.

\subsection{An Alternative Formulation of the Attractor Equations}

The flux attractor equations \eqref{eq:FullAttEqsC1}-\eqref{eq:FullAttEqsC4}
contain Kähler covariant derivatives, which we find much less convenient
than ordinary derivatives. We therefore consider a modified version
of \eqref{eq:HelloC^i} that does not have this problem:\begin{equation}
G_{3}=-i\mbox{Im}\left(\tau\right)\left[\overline{C}\overline{\Omega}_{3}+L^{I}\partial_{I}\Omega_{3}\right],\label{eq:HelloL^I}\end{equation}
where $\overline{C}$ and the $L^{I}$ are coefficients. Note that
we differentiate with respect to the $Z^{I},$ not the $z^{i}.$ 

The ISD condition \eqref{eq:ISD} allows only $\left(2,1\right)$
and $\left(0,3\right)$ pieces in the complex flux $G_{3}.$ While
the \emph{ansatz} \eqref{eq:HelloL^I} does not contain a $\left(1,2\right)$
piece, equation \eqref{eq:dOmega} shows that the $\partial_{I}\Omega_{3}$
term includes a $\left(3,0\right)$ piece. Since the ISD condition
\eqref{eq:ISD} forbids such a term, we must choose the $L^{I}$ so
that it is projected out. The appropriate condition on the $L^{I}$
is \begin{equation}
L^{I}\partial_{I}K_{z}=0\,.\label{eq:Hello-Constraint}\end{equation}
After imposing this condition, the resulting $G_{3}$ has only $\left(0,3\right)$
and $\left(2,1\right)$ pieces. We thus conclude that \eqref{eq:HelloL^I}
and \eqref{eq:Hello-Constraint} together are equivalent to \eqref{eq:HelloC^i},
with\begin{equation}
C^{i}=\frac{\partial z^{i}}{\partial Z^{I}}L^{I}\,.\label{eq:CtoL}\end{equation}
If we think of the $C^{i}$ as given, then this fixes $n$ of the
$n+1$ components of $L^{I},$ and \eqref{eq:Hello-Constraint} fixes
the final component. 

As in section \ref{sub:Flux-Attractor-Equations}, we can expand \eqref{eq:HelloL^I}
and find a set of real attractor equations. This is equivalent to
replacing $C^{i}D_{i}\to L^{I}\partial_{I}$ in \eqref{eq:FullAttEqsC1}-\eqref{eq:FullAttEqsC4}
and adding the constraint \eqref{eq:Hello-Constraint}. The resulting
attractor equations are:\begin{eqnarray}
m_{h}^{I} & = & \mbox{Re}\left[CZ^{I}+L^{I}\right],\label{eq:dAtt-1}\\
m_{f}^{I} & = & \mbox{Re}\left[\tau CZ^{I}+\overline{\tau}L^{I}\right],\label{eq:dAtt-2}\\
e_{I}^{h} & = & \mbox{Re}\left[CF_{I}+L^{J}F_{IJ}\right],\label{eq:dAtt-3}\\
e_{I}^{f} & = & \mbox{Re}\left[\tau CF_{I}+\overline{\tau}L^{J}F_{IJ}\right],\label{eq:dAtt-4}\\
0 & = & L^{I}\left(\overline{F}_{I}-\overline{Z}^{J}F_{IJ}\right),\label{eq:3-0-Constraint}\end{eqnarray}
where we have introduced $F_{IJ}\equiv\partial_{I}F_{J},$ and used
\eqref{eq:Oab} to make the constraint \eqref{eq:Hello-Constraint}
more explicit. The magnetic attractor equations \eqref{eq:dAtt-1}
and \eqref{eq:dAtt-2} are simpler than their counterparts \eqref{eq:FullAttEqsC1}
and \eqref{eq:FullAttEqsC2}, in that the $C^{i}D_{i}Z^{I}$ term
reduces to $L^{I}.$ Similarly, the electric attractor equations \eqref{eq:dAtt-3}
and \eqref{eq:dAtt-4} are simpler than \eqref{eq:FullAttEqsC3} and
\eqref{eq:FullAttEqsC4} since the Kähler covariant derivatives have
been replaced with ordinary derivatives. 

Another benefit of these reformulated attractor equations is that
the $L^{I}$ transform in the $n+1$ of $SO\left(n+1\right),$ just
like the $Z^{I}$ and the fluxes, and in contrast to the $C^{i}.$
This suggests solving \eqref{eq:dAtt-1}-\eqref{eq:dAtt-4} for $CZ^{I}$
and $L^{I},$ treating the $L^{I}$ on an equal footing with the $CZ^{I},$
then solving \eqref{eq:3-0-Constraint} for $\tau.$ This procedure
is more practical than solving \eqref{eq:FullAttEqsC1}-\eqref{eq:FullAttEqsC4}
for the $n+1$ vector $CZ^{I},$ $n$ vector $C^{i},$ and scalar
$\tau,$ even though the results are equivalent. We will demonstrate
this by completely solving an explicit example in section \ref{sec:STUSolution}.

\subsection{Magnetic Attractor Equations and the Mixed Ensemble}

We now solve the flux attractor equations by adapting the OSV procedure
for solving the black hole attractor equations \citep{Ooguri:2004zv}.
We treat $\tau$ as a fixed variable while solving \eqref{eq:dAtt-1}-\eqref{eq:dAtt-4},
then determine it at the very end by solving \eqref{eq:3-0-Constraint}.
The two sets of variables we have seen so far, $\left\{ CZ^{I},L^{I},\tau\right\} $
and $\left\{ m_{h}^{I},m_{f}^{I},e_{I}^{h},e_{I}^{f},\tau\right\} ,$
describe two different ensembles. Following OSV, we introduce a {}``mixed
ensemble,'' $\left\{ m_{h}^{I},m_{f}^{I},\phi_{h}^{I},\phi_{f}^{I},\tau\right\} ,$
where $\phi_{h,f}^{I}$ are potentials conjugate to the electric fluxes.
When introducing these potentials, we require that:
\begin{enumerate}
\item The expressions for $CZ^{I}$ and $L^{I}$ in terms of $\left\{ m_{h}^{I},m_{f}^{I},\phi_{h}^{I},\phi_{f}^{I},\tau\right\} $
automatically solve the {}``magnetic'' attractor equations, \eqref{eq:dAtt-1}
and \eqref{eq:dAtt-2}.
\item The potentials $\left\{ \phi_{h}^{I},\phi_{f}^{I}\right\} $ transform
like $\left\{ m_{h}^{I},m_{f}^{I}\right\} $ under S-duality.
\item The relationship between $\left\{ CZ^{I},L^{I},\tau\right\} $ and
$\left\{ m_{h}^{I},m_{f}^{I},\phi_{h}^{I},\phi_{f}^{I},\tau\right\} $
is covariant under S-duality.
\end{enumerate}
These conditions determine the relationship between $\left\{ CZ^{I},L^{I},\tau\right\} $
and $\left\{ m_{h}^{I},m_{f}^{I},\phi_{h}^{I},\phi_{f}^{I},\tau\right\} $
to be\begin{eqnarray}
CZ^{I} & = & \frac{1}{\tau-\overline{\tau}}\left(m_{f}^{I}-\overline{\tau}m_{h}^{I}\right)+\frac{1}{\tau-\overline{\tau}}\left(\phi_{f}^{I}-\overline{\tau}\phi_{h}^{I}\right),\label{eq:Z-expr}\\
L^{I} & = & -\frac{1}{\tau-\overline{\tau}}\left(m_{f}^{I}-\tau m_{h}^{I}\right)+\frac{1}{\tau-\overline{\tau}}\left(\phi_{f}^{I}-\tau\phi_{h}^{I}\right).\label{eq:L-expr}\end{eqnarray}
We will also want to know how derivatives with respect to $Z^{I}$
and $L^{I}$ are mapped into derivatives with respect to fluxes and
the potentials. Here it is important to note that both sets of variables
we are considering, $\left\{ CZ^{I},L^{I},\tau\right\} $ and $\left\{ m_{h}^{I},m_{f}^{I},\phi_{h}^{I},\phi_{f}^{I},\tau\right\} ,$
include $\tau$ as an \emph{independent} variable. The derivatives
are therefore related by\begin{eqnarray}
\frac{1}{C}\frac{\partial}{\partial Z^{I}} & = & \frac{1}{2}\left[\left(\frac{\partial}{\partial m_{h}^{I}}+\tau\frac{\partial}{\partial m_{f}^{I}}\right)+\left(\frac{\partial}{\partial\phi_{h}^{I}}+\tau\frac{\partial}{\partial\phi_{f}^{I}}\right)\right],\label{eq:Z-deriv}\\
\frac{\partial}{\partial L^{I}} & = & \frac{1}{2}\left[\left(\frac{\partial}{\partial m_{h}^{I}}+\overline{\tau}\frac{\partial}{\partial m_{f}^{I}}\right)-\left(\frac{\partial}{\partial\phi_{h}^{I}}+\overline{\tau}\frac{\partial}{\partial\phi_{f}^{I}}\right)\right],\label{eq:L-deriv}\end{eqnarray}
where all derivatives are taken with $\tau$ held fixed.

\subsection{\label{sub:Electric-AttEqs-Gen}Electric Attractor Equations and
the Generating Function}

In the previous section we solved the magnetic attractor equations,
\eqref{eq:dAtt-1} and \eqref{eq:dAtt-2}. We now introduce an auxiliary
function, \begin{equation}
\mathcal{V}=2\mbox{Im}\left(\tau\right)CF_{I}L^{I}\,,\label{eq:V-def}\end{equation}
that simplifies the electric attractor equations, \eqref{eq:dAtt-3}
and \eqref{eq:dAtt-4}. This new function plays a role analogous to
that of the prepotential in the solution of the black hole attractor
equations. It enjoys the following properties:
\begin{enumerate}
\item Derivatives of $\mathcal{V}$ with respect to $L^{I}$ give $CF_{I},$
one of the terms that appears in the electric attractor equations:\begin{equation}
\frac{1}{2\mbox{Im}\left(\tau\right)}\frac{\partial\mathcal{V}}{\partial L^{I}}=CF_{I}\,.\end{equation}

\item Derivatives with respect to $Z^{I}$ give $L^{J}F_{IJ},$ the other
term that appears in the electric attractor equations:\begin{equation}
\frac{1}{2C\mbox{Im}\left(\tau\right)}\frac{\partial\mathcal{V}}{\partial Z^{I}}=L^{J}F_{IJ}\,.\end{equation}

\item The factor of $C$ in \eqref{eq:V-def}makes $\mathcal{V}$ invariant
under Kähler transformations.
\item By \eqref{eq:Coeffs-S-tform}, \eqref{eq:Coeffs-S-tform-2}, and \eqref{eq:Imtau-S-tform},
the factor of $\mbox{Im}\left(\tau\right)$ in \eqref{eq:V-def} makes
$\mathcal{V}$ invariant under S-duality.
\item $\mathcal{V}$ is holomorphic in $L^{I}$ and $Z^{I}.$
\end{enumerate}
The first two properties will allow us to replace the $F_{I}$ and
$L^{J}F_{IJ}$ terms in the electric attractor equations, \eqref{eq:dAtt-3}
and \eqref{eq:dAtt-4}, with derivatives of $\mathcal{V}.$ This is
analogous to the role played by the prepotential in the solution of
the electric black hole equations. The invariance of $\mathcal{V}$
under Kähler transformations and S-duality (properties 3 and 4) will
allow us to interpret it in terms of a physical quantity. Finally,
we will make extensive use of holomorphy in the following manipulations.

As described above, we can rewrite the electric attractor equations
\eqref{eq:dAtt-3} and \eqref{eq:dAtt-4} in terms of derivatives
of $\mathcal{V},$\begin{eqnarray}
e_{I}^{h} & = & \frac{1}{2\mbox{Im}\left(\tau\right)}\mbox{Re}\left[\left.\frac{\partial\mathcal{V}}{\partial L^{I}}\right|_{Z^{J},L^{J\neq I},\tau}+\frac{1}{C}\left.\frac{\partial\mathcal{V}}{\partial Z^{I}}\right|_{Z^{J\neq I},L^{J},\tau}\right],\\
e_{I}^{f} & = & \frac{1}{2\mbox{Im}\left(\tau\right)}\mbox{Re}\left[\tau\left.\frac{\partial\mathcal{V}}{\partial L^{I}}\right|_{Z^{J},L^{J\neq I},\tau}+\overline{\tau}\frac{1}{C}\left.\frac{\partial\mathcal{V}}{\partial Z^{I}}\right|_{Z^{J\neq I},L^{J},\tau}\right].\end{eqnarray}
We then use holomorphy of $\mathcal{V}$ to find\begin{eqnarray}
e_{I}^{h} & = & \frac{i}{2\mbox{Im}\left(\tau\right)}\left(\frac{\partial}{\partial L^{I}}+\frac{1}{C}\cdot\frac{\partial}{\partial Z^{I}}-\frac{\partial}{\partial\overline{L}^{I}}-\frac{1}{\overline{C}}\frac{\partial}{\partial\overline{Z}^{I}}\right)\mbox{Im}\left(\mathcal{V}\right),\\
e_{I}^{f} & = & \frac{i}{2\mbox{Im}\left(\tau\right)}\left\{ \tau\left(\frac{\partial}{\partial L^{I}}-\frac{1}{\overline{C}}\cdot\frac{\partial}{\partial\overline{Z}^{I}}\right)-\overline{\tau}\left(\frac{\partial}{\partial\overline{L}^{I}}-\frac{1}{C}\frac{\partial}{\partial Z^{I}}\right)\right\} \mbox{Im}\left(\mathcal{V}\right).\end{eqnarray}
Finally, we introduce derivatives with respect to the potentials using
\eqref{eq:Z-deriv} and \eqref{eq:L-deriv},\begin{eqnarray}
e_{I}^{h} & = & -\left[\frac{\partial}{\partial\phi_{f}^{I}}\mbox{Im}\left(\mathcal{\mathcal{V}}\right)\right]_{\phi_{h}^{J\neq I},\phi_{f}^{J},m_{h}^{J},m_{f}^{J},\tau},\label{eq:eih-tau}\\
e_{I}^{f} & = & \left[\frac{\partial}{\partial\phi_{h}^{I}}\mbox{Im}\left(\mathcal{\mathcal{V}}\right)\right]_{\phi_{h}^{J},\phi_{f}^{J\neq I},m_{h}^{J},m_{f}^{J},\tau},\label{eq:eif-tau}\end{eqnarray}
Though we initially defined $\mathcal{V}$ in terms of $L^{I}$ and
$Z^{I},$ in this last step we simply substitute in \eqref{eq:Z-expr}
and \eqref{eq:L-expr} to make it a function of the magnetic fluxes
and electric potentials. 

It is remarkable that the electric attractor equations, which appear
rather complex, reduce to derivatives of a single generating function!
This is one of the principal results of this paper. 

Since we have made a rather long chain of substitutions and redefinitions,
we briefly summarize our procedure for solving the flux attractor
equations:
\begin{enumerate}
\item \label{enu:Inputs}Take as inputs the fluxes $\left\{ m_{f}^{I},m_{h}^{I},e_{I}^{f},e_{I}^{h}\right\} $
and the symplectic section $\left\{ Z^{I},F_{I}\right\} .$
\item Insert the expressions for the $F_{I}$ as functions of the $Z^{I}$
into \eqref{eq:V-def}, giving $\mathcal{V}\left(L^{I},CZ^{I},\tau\right).$ 
\item Substitute the expressions \eqref{eq:Z-expr} and \eqref{eq:L-expr}
into $\mathcal{V}\left(L^{I},CZ^{I},\tau\right)$ to get $\mathcal{V}\left(\phi_{h}^{I},\phi_{f}^{I},m_{h}^{I},m_{f}^{I},\tau\right).$
\item \label{enu:Invert}Invert \eqref{eq:eih-tau} and \eqref{eq:eif-tau}
to get expressions for $\phi_{f}^{I}$ and $\phi_{h}^{I}$ in terms
of $m_{h}^{I},$ $m_{f}^{I},$ $e_{I}^{h},$ $e_{I}^{f},$ and $\tau.$ 
\item Rewrite the constraint \eqref{eq:3-0-Constraint} in terms of $m_{h}^{I},$
$m_{f}^{I},$ $e_{I}^{h},$ $e_{I}^{f},$ and $\tau.$ Do this by
substituting \eqref{eq:Z-expr} and \eqref{eq:L-expr} into \eqref{eq:3-0-Constraint},
then inserting the solutions for $\phi_{f}^{I}$ and $\phi_{h}^{I}$
in terms of $m_{h}^{I},$ $m_{f}^{I},$ $e_{I}^{h},$ $e_{I}^{f},$
and $\tau.$
\item Solve the constraint \eqref{eq:3-0-Constraint} for $\tau$ as a function
of the fluxes only. Substitute this back into the expressions for
$\phi_{f,h}^{I}$ to get expressions for the potentials in terms of
the fluxes only, and then insert $\tau$ and the potentials into the
expressions \eqref{eq:Z-expr} and \eqref{eq:L-expr} to get expressions
for $CZ^{I}$ and $L^{I}$ in terms of the fluxes only.
\end{enumerate}
The most difficult part of this procedure is step \ref{enu:Invert},
which requires that we invert a system of $2n+2$ equations. Even
in simple cases, these result in polynomials of impractically high
order. 

The electric attractor equations \eqref{eq:eih-tau} and \eqref{eq:eif-tau}
take the form of thermodynamic relations, indicating that the potentials
$\phi_{f,h}^{I}$ are \emph{conjugate} to the fluxes $e_{I}^{f,h}.$
This suggests the Legendre transformation\begin{equation}
\mathcal{G}=\mbox{Im}\left(\mathcal{V}\right)+e_{I}^{h}\phi_{f}^{I}-e_{I}^{f}\phi_{h}^{I}\,,\label{eq:G-def}\end{equation}
so that the electric attractor equations become\begin{eqnarray}
\phi_{h}^{I} & = & -\left[\frac{\partial\mathcal{G}}{\partial e_{I}^{f}}\right]_{e_{J}^{h},e_{J\neq I}^{f},m_{h}^{J},m_{f}^{J},\tau},\label{eq:phih-from-G}\\
\phi_{f}^{I} & = & \left[\frac{\partial\mathcal{G}}{\partial e_{I}^{h}}\right]_{e_{J\neq I}^{h},e_{J}^{f},m_{h}^{J},m_{f}^{J},\tau}.\label{eq:phif-from-G}\end{eqnarray}
This means that we only need to know a single function, $\mathcal{G},$
which is in principle determined by steps \ref{enu:Inputs}-\ref{enu:Invert}
above. 

In practice, this may not be the best way to proceed. The analogue
of $\mathcal{G}$ for the black hole attractor equations is the entropy
$S$, which can be computed by many different methods. For example,
the requirement that $S$ be invariant under duality transformations
severely constrains, and sometimes completely determines, its functional
form \citep{Kallosh:1996uy}.

\subsection{The Constraint and the Generating Function}

So far, we have demonstrated that the electric attractor equations
\eqref{eq:dAtt-3} and \eqref{eq:dAtt-4} can be recast in terms of
derivatives of a generating function. Indeed, we designed the generating
function $\mathcal{G}$ specifically for this purpose. Next, we demonstrate
a more surprising result: the constraint \eqref{eq:3-0-Constraint}
can \emph{also} be written in terms of derivatives of the same generating
function. 

We first compute $\tau-$derivatives of $CZ^{I}$ and $L^{I}$ in
the $\left\{ m_{h}^{I},m_{f}^{I},\phi_{h}^{I},\phi_{f}^{I},\tau\right\} $
ensemble, using \eqref{eq:Z-expr} and \eqref{eq:L-deriv}:

\begin{eqnarray}
\left.\frac{\partial Z^{I}}{\partial\tau}\right|_{m_{h}^{I},m_{f}^{I},\phi_{h}^{I},\phi_{f}^{I}}=-\frac{Z^{I}}{\tau-\overline{\tau}}\,, &  & \left.\frac{\partial\overline{Z}^{I}}{\partial\tau}\right|_{m_{h}^{I},m_{f}^{I},\phi_{h}^{I},\phi_{f}^{I}}=\frac{Z^{I}}{\tau-\overline{\tau}}\,,\\
\left.\frac{\partial L^{I}}{\partial\tau}\right|_{m_{h}^{I},m_{f}^{I},\phi_{h}^{I},\phi_{f}^{I}}=\frac{\overline{L}^{I}}{\tau-\overline{\tau}}\,, &  & \left.\frac{\partial\overline{L}^{I}}{\partial\tau}\right|_{m_{h}^{I},m_{f}^{I},\phi_{h}^{I},\phi_{f}^{I}}=-\frac{\overline{L}^{I}}{\tau-\overline{\tau}}\,.\end{eqnarray}
Using these preliminary results, we find:\begin{eqnarray}
\frac{\partial}{\partial\tau}\left[\mbox{Im}\left(\mathcal{V}\right)\right]_{m,\phi} & = & \frac{\partial}{\partial\tau}\left[2\mbox{Im}\left(\tau\right)\mbox{Im}\left(L^{I}F_{I}\right)\right]_{m,\phi}\\
 & = & -i\mbox{Im}\left(L^{I}F_{I}\right)-i\mbox{Im}\left(\tau\right)\left[\frac{\overline{L}^{I}}{\tau-\overline{\tau}}F_{I}+\frac{\overline{L}^{I}}{\tau-\overline{\tau}}\overline{F}_{I}\right]\nonumber \\
 &  & -i\mbox{Im}\left(\tau\right)\left[-L^{I}F_{IJ}\frac{Z^{J}}{\tau-\overline{\tau}}-\overline{L}^{I}\overline{F}_{IJ}\frac{Z^{J}}{\tau-\overline{\tau}}\right]\\
 & = & \frac{1}{2}\left[-L^{I}F_{I}-\overline{L}^{I}F_{I}+L^{I}F_{IJ}Z^{J}+\overline{L}^{J}\overline{F}_{IJ}Z^{J}\right]\\
 & = & -\frac{1}{2}\overline{L}^{I}\left[F_{I}-\overline{F}_{IJ}Z^{J}\right],\label{eq:ConstraintRecovered}\end{eqnarray}
using the homogeneity property $F_{IJ}Z^{J}=F_{I}.$ The last line
is proportional to the complex conjugate of the constraint \eqref{eq:3-0-Constraint}.
Setting $\partial\mbox{Im}\left(\mathcal{V}\right)/\partial\tau=0$
is thus equivalent to imposing \eqref{eq:3-0-Constraint}. Notice
that the overall factor of $\mbox{Im}\left(\tau\right)$ included
in $\mathcal{V},$ originally introduced to make $\mathcal{V}$ invariant
under S-duality, is exactly what is required to recover the constraint
\eqref{eq:3-0-Constraint} from $\partial\mbox{Im}\left(\mathcal{V}\right)/\partial\tau.$

The Legendre transform that takes us from the $\left\{ \phi_{h}^{I},\phi_{f}^{I},m_{h}^{I},m_{f}^{I},\tau\right\} $
ensemble to the $\left\{ e_{I}^{h},e_{I}^{f},m_{h}^{I},m_{f}^{I},\tau\right\} $
ensemble does not change the equilibrium condition associated with
$\tau.$ In the latter ensemble, the constraint \eqref{eq:3-0-Constraint}
is equivalent to\begin{equation}
\left.\frac{\partial\mathcal{G}}{\partial\tau}\right|_{e_{I}^{h},e_{I}^{f},m_{h}^{I},m_{f}^{I}}=0.\label{eq:ConstraintFromG}\end{equation}
This completes our demonstration that the flux attractor equations
can be interpreted as equilibrium conditions for a thermodynamic system.
From the thermodynamic point of view, \eqref{eq:ConstraintRecovered}
indicates that $\tau$ is \emph{conjugate} to the constraint \eqref{eq:3-0-Constraint}. 

While studying $\mathcal{G}$ in the $\left\{ e_{I}^{h},e_{I}^{f},m_{h}^{I},m_{f}^{I},\tau\right\} $
ensemble may be conceptually clearer, there is a useful consequence
of \eqref{eq:ConstraintFromG}. Suppose we take derivatives of $\mathcal{G}$
\emph{without} holding $\tau$ fixed. The result is: \begin{eqnarray}
\left.\frac{\partial\mathcal{G}}{\partial e_{I}^{h}}\right|_{e_{J\neq I}^{h},e_{J}^{f},m_{h}^{J},m_{f}^{J}} & = & \left.\frac{\partial\mathcal{G}}{\partial e_{I}^{h}}\right|_{e_{J\neq I}^{h},e_{J}^{f},m_{h}^{J},m_{f}^{J},\tau}+\left.\frac{\partial\mathcal{G}}{\partial\tau}\right|_{e_{I}^{h},e_{I}^{f},m_{h}^{I},m_{f}^{I}}\left.\frac{\partial\tau}{\partial e_{I}^{h}}\right|_{e_{J\neq I}^{h},e_{J}^{f},m_{h}^{J},m_{f}^{J}}\\
 & = & \left.\frac{\partial\mathcal{G}}{\partial e_{I}^{h}}\right|_{e_{J\neq I}^{h},e_{J}^{f},m_{h}^{J},m_{f}^{J},\tau}.\end{eqnarray}
In other words, if we substitute the attractor value for $\tau$ into
$\mathcal{G}$ we can simplify \eqref{eq:phih-from-G} and \eqref{eq:phif-from-G}
to:\begin{eqnarray}
\phi_{h}^{I} & = & -\left[\frac{\partial\mathcal{G}}{\partial e_{I}^{f}}\right]_{e_{J}^{h},e_{J\neq I}^{f},m_{h}^{J},m_{f}^{J}},\label{eq:phih-G-notau}\\
\phi_{f}^{I} & = & \left[\frac{\partial\mathcal{G}}{\partial e_{I}^{h}}\right]_{e_{J\neq I}^{h},e_{J}^{f},m_{h}^{J},m_{f}^{J}}.\label{eq:phif-G-notau}\end{eqnarray}
If one can determine $\mathcal{G}$ as a function of arbitrary fluxes,
then \eqref{eq:phih-G-notau} and \eqref{eq:phif-G-notau} determine
the potentials $\phi_{h,f}^{I},$ \eqref{eq:Z-expr} and \eqref{eq:L-expr}
then determine the moduli $Z^{I}$ and mass parameters $L^{I},$ and
finally \eqref{eq:3-0-Constraint} determines the axio-dilaton $\tau.$
In this way the single function $\mathcal{G}$ determines the vacuum
expectation values and masses of the moduli.

\section{\label{sec:GeneralPropertiesOfG}General Properties of the Generating
Function}

The generating function $\mathcal{G}$ introduced in \eqref{eq:G-def}
is the function that controls the flux attractor, giving attractor
values for scalars and other physical quantities upon differentiation.
In this section we initiate a general study of the generating function
by demonstrating a simple relationship between $\mathcal{G}$ and
the gravitino mass:\begin{equation}
\mathcal{G}=\int F_{3}\wedge H_{3}-2\mbox{Vol}^{2}m_{3/2}^{2}\,.\label{eq:GeneralG}\end{equation}
 Note that the gravitino mass is to be considered a function of arbitrary
fluxes. We first introduce a condensed, complex notation for the fluxes
and potentials. We then exploit the homogeneity properties of $\mathcal{G}$
to prove the relationship \eqref{eq:GeneralG}.

\subsection{Complex Fluxes and Potentials}

One of the results of section \ref{sec:Solving-the-Flux} is that
we can solve the electric and magnetic attractor equations \eqref{eq:dAtt-1}-\eqref{eq:dAtt-4}
treating $\tau$ as a constant, then determine $\tau$ by solving
\eqref{eq:3-0-Constraint}. This justifies the introduction of the
following complex fluxes and potentials:\begin{eqnarray}
m^{I} & \equiv & m_{f}^{I}-\tau m_{h}^{I}\,,\label{eq:ComplexMs}\\
e_{I} & \equiv & e_{I}^{f}-\tau e_{I}^{h}\,,\label{eq:ComplexEs}\\
\varphi^{I} & \equiv & \phi_{f}^{I}-\tau\phi_{h}^{I}\,.\label{eq:ComplexPhis}\end{eqnarray}
We can then use \eqref{eq:ComplexMs} and \eqref{eq:ComplexPhis}
to rewrite \eqref{eq:Z-expr} and \eqref{eq:L-expr} as\begin{eqnarray}
CZ^{I} & = & \frac{1}{2i\mbox{Im}\left(\tau\right)}\left[\overline{m}^{I}+\overline{\varphi}^{I}\right],\label{eq:CZI-complex}\\
L^{I} & = & \frac{1}{2i\mbox{Im}\left(\tau\right)}\left[-m^{I}+\varphi^{I}\right].\label{eq:LI-complex}\end{eqnarray}
We also define derivatives with respect to the complex electric fluxes
as\begin{equation}
\frac{\partial}{\partial e_{I}}\equiv\frac{i}{2\mbox{Im}\left(\tau\right)}\left(\frac{\partial}{\partial e_{I}^{h}}+\overline{\tau}\frac{\partial}{\partial e_{I}^{f}}\right),\label{eq:e-deriv-Complex}\end{equation}
where the normalization is chosen so that $\partial e_{I}/\partial e_{J}=\delta_{I}^{J}$.
Definitions for $\partial/\partial m^{I}$ and $\partial/\partial\varphi^{I}$
are completely analogous. We can then rewrite the electric attractor
equations \eqref{eq:phih-G-notau} and \eqref{eq:phif-G-notau} as\begin{equation}
\varphi^{I}=2i\mbox{Im}\left(\tau\right)\frac{\partial\mathcal{G}}{\partial\overline{e}^{I}}\,,\end{equation}
and the expressions for $CZ^{I}$ and $L^{I}$ as\begin{eqnarray}
CZ^{I} & = & \frac{1}{2i\mbox{Im}\left(\tau\right)}\left[\overline{m}^{I}-2i\mbox{Im}\left(\tau\right)\frac{\partial}{\partial e_{I}}\mathcal{G}\right],\label{eq:CZI-again}\\
L^{I} & = & \frac{1}{2i\mbox{Im}\left(\tau\right)}\left[-m^{I}+2i\mbox{Im}\left(\tau\right)\frac{\partial}{\partial\overline{e}_{I}}\mathcal{G}\right].\label{eq:LI-again}\end{eqnarray}

While \eqref{eq:CZI-again} and \eqref{eq:LI-again} present a fairly
compact version of the results of section \eqref{sec:Solving-the-Flux},
they treat the electric and magnetic fluxes quite differently. The
generating function $\mathcal{G}$ is not homogeneous in either the
electric or the magnetic fluxes \emph{alone},\emph{ }so a symplectic
invariant version of \eqref{eq:CZI-again} and \eqref{eq:LI-again}
will be helpful. We formulate this by first introducing a new operator:\begin{eqnarray}
\partial & \equiv & \alpha_{I}\frac{\partial}{\partial e_{I}}+\beta^{I}\frac{\partial}{\partial m^{I}}\,,\label{eq:DiffOp}\end{eqnarray}
which maps scalar functions of the fluxes to 3-forms. We then examine
\eqref{eq:dAtt-1}-\eqref{eq:dAtt-4}, and see that symplectic invariance
requires \begin{eqnarray}
C\Omega_{3} & = & \frac{1}{2i\mbox{Im}\left(\tau\right)}\left[\overline{G}_{3}-2i\mbox{Im}\left(\tau\right)\partial\mathcal{G}\right],\label{eq:Att-O}\\
L^{I}\partial_{I}\Omega_{3} & = & \frac{1}{2i\mbox{Im}\left(\tau\right)}\left[-G_{3}+2i\mbox{Im}\left(\tau\right)\overline{\partial}\mathcal{G}\right].\label{eq:Att-L}\end{eqnarray}
These are equivalent to the electric attractor equations, so they
must be supplemented by the constraint \eqref{eq:3-0-Constraint}.
This amounts to some flexibility in our treatment of $\mathcal{G}.$
We can either use $\mathcal{G}\left(e_{I},m^{I},\tau\right)$ and
take all derivatives with $\tau$ held fixed, as in \eqref{eq:phih-from-G}
and \eqref{eq:phif-from-G}, or substitute in the attractor value
of $\tau$ to find $\mathcal{G}\left(e_{I},m^{I}\right)$ and differentiate
as in \eqref{eq:phih-G-notau} and \eqref{eq:phif-G-notau}.

\subsection{\label{sub:Gen-G-Gravitino}General Expression for the Generating
Function}

We now show that the relationship between the generating function
$\mathcal{G}$ and the gravitino mass \eqref{eq:GeneralG} holds for
general compactifications. Our argument turns on a homogeneity property
of the attractor equations that is evident from examining \eqref{eq:dAtt-1}-\eqref{eq:3-0-Constraint}.
These attractor equations are invariant under a uniform rescaling
of the fluxes,\begin{eqnarray}
m_{h,f}^{I} & \to & e^{\lambda}m_{h,f}^{I}\,,\\
e_{I}^{h,f} & \to & e^{\lambda}e_{I}^{h,f}\,,\end{eqnarray}
provided that we simultaneously rescale\begin{eqnarray}
CZ^{I} & \to & e^{\lambda}CZ^{I}\,,\\
L^{I} & \to & e^{\lambda}L^{I}\,.\end{eqnarray}
If we then turn our attention to the expressions for the $CZ^{I}$
and $L^{I}$ in terms of fluxes and potentials,\eqref{eq:Z-expr}
and \eqref{eq:L-expr}, we see that the potentials must transform
as \begin{equation}
\phi_{h,f}^{I}\to e^{\lambda}\phi_{h,f}^{I}\,.\end{equation}
Equations \eqref{eq:phih-G-notau} and \eqref{eq:phif-G-notau} then
indicate that if the potentials are to be homogeneous of degree one
in the fluxes, then $\mathcal{G}$ must be homogeneous of degree two
in the fluxes. If we use the complex fluxes introduced in \eqref{eq:ComplexMs}
and \eqref{eq:ComplexEs}, we find that $\mathcal{G}$ is homogeneous
of degree one in the complex fluxes and degree one in their conjugates.
This homogeneity implies that\begin{equation}
\int G_{3}\wedge\partial\mathcal{G}=\left[m^{I}\frac{\partial}{\partial m^{I}}+e_{I}\frac{\partial}{\partial e_{I}}\right]\mathcal{G}=\mathcal{G},\label{eq:G-dCalG}\end{equation}
where we used the orthogonality relations \eqref{eq:Ortho1} and \eqref{eq:Ortho2}
and expansions \eqref{eq:Hab} and \eqref{eq:Fab} to compute the
integral. We will now use this result to compute the superpotential
and Kähler potential at the attractor point, and finally the gravitino
mass.

We begin with the superpotential \eqref{eq:GVW}, then substitute
in \eqref{eq:Att-O}:\begin{eqnarray}
CW & = & \int G_{3}\wedge C\Omega_{3}\\
 & = & \frac{1}{2i\mbox{Im}\left(\tau\right)}\left[\int G_{3}\wedge\overline{G}_{3}-2i\mbox{Im}\left(\tau\right)\int G_{3}\wedge\partial\mathcal{G}\right]\\
 & = & \int F_{3}\wedge H_{3}-\mathcal{G}.\end{eqnarray}
In order to determine the Kähler potential we need to compute\begin{eqnarray}
\left|C\right|^{2}\int\Omega_{3}\wedge\overline{\Omega}_{3} & = & \frac{1}{4\mbox{Im}\left(\tau\right)^{2}}\int\left(\overline{G}_{3}-2i\mbox{Im}\left(\tau\right)\partial\mathcal{G}\right)\wedge\left(G_{3}+2i\mbox{Im}\left(\tau\right)\overline{\partial}\mathcal{G}\right)\\
 & = & \frac{1}{4\mbox{Im}\left(\tau\right)^{2}}\left[-\int G_{3}\wedge\overline{G}_{3}+2i\mbox{Im}\left(\tau\right)\left(\int G_{3}\wedge\partial\mathcal{G}+\int\overline{G}_{3}\wedge\overline{\partial}\mathcal{G}\right)\right.\nonumber \\
 &  & \left.+4\mbox{Im}\left(\tau\right)^{2}\int\partial\mathcal{G}\wedge\overline{\partial}\mathcal{G}\right]\\
 & = & -\frac{i}{\mbox{Im}\left(\tau\right)}\left[\int F_{3}\wedge H_{3}-\mathcal{G}\right].\end{eqnarray}
In the last step we used \eqref{eq:G-dCalG} and \begin{equation}
4\mbox{Im}\left(\tau\right)^{2}\int\partial\mathcal{G}\wedge\overline{\partial}\mathcal{G}=-\int G_{3}\wedge\overline{G}_{3}\,,\label{eq:dg-dbarg}\end{equation}
which we prove as follows. $L^{I}\partial_{I}\Omega_{3}$ contains
only $\left(3,0\right)$ and $\left(2,1\right)$ pieces, so if we
integrate it against $\Omega_{3}$ the result must vanish: \begin{eqnarray}
0 & = & \int C\Omega_{3}\wedge L^{I}\partial_{I}\Omega_{3}\\
 & = & -\frac{1}{4\mbox{Im}\left(\tau\right)^{2}}\int\left(\overline{G}_{3}-2i\mbox{Im}\left(\tau\right)\partial\mathcal{G}\right)\wedge\left(-G_{3}+2i\mbox{Im}\left(\tau\right)\overline{\partial}\mathcal{G}\right)\\
 & = & -\frac{1}{4\mbox{Im}\left(\tau\right)^{2}}\left[\int G_{3}\wedge\overline{G}_{3}+2i\mbox{Im}\left(\tau\right)\left(\int\overline{G}_{3}\wedge\overline{\partial}\mathcal{G}-\int G_{3}\wedge\partial\mathcal{G}\right)\right.\nonumber \\
 &  & \left.+4\mbox{Im}\left(\tau\right)^{2}\int\partial\mathcal{G}\wedge\overline{\partial}\mathcal{G}\right]\\
 & = & -\frac{1}{4\mbox{Im}\left(\tau\right)^{2}}\left[\int G_{3}\wedge\overline{G}_{3}+4\mbox{Im}\left(\tau\right)^{2}\int\partial\mathcal{G}\wedge\overline{\partial}\mathcal{G}\right],\end{eqnarray}
which implies \eqref{eq:dg-dbarg}. 

We now write out the gravitino mass \eqref{eq:GraivitinoDef} with
the full Kähler potential \eqref{eq:KahlerPot}:\begin{eqnarray}
\mbox{Vol}^{2}m_{3/2}^{2} & = & \frac{\left|CW\right|^{2}}{2i\mbox{Im}\left(\tau\right)\left|C\right|^{2}\int\Omega_{3}\wedge\overline{\Omega}_{3}}\\
 & = & \frac{1}{2}\left[\int F_{3}\wedge H_{3}-\mathcal{G}\right].\end{eqnarray}
Reorganizing this we find the generating function,\begin{equation}
\mathcal{G}=\int F_{3}\wedge H_{3}-2\mbox{Vol}^{2}m_{3/2}^{2}\,,\end{equation}
as we wanted to show. We also point out a curious relationship:\begin{equation}
\mbox{Vol}^{2}m_{3/2}^{2}=\frac{1}{2}CW\,,\end{equation}
where both quantities are evaluated at the attractor point. One could
have imagined that other duality-invariant quantities, e.g. eigenvalues
of the mass matrix, would appear in one or more of these expressions,
but they do not. We also point out that the combination $\mbox{Vol}^{2}m_{3/2}^{2}$
is independent of the Kähler moduli, which cannot be stabilized by
turning on 3-form fluxes.

As a side product of our derivation, we find another interesting identity.
While one combination of \eqref{eq:Att-O} and \eqref{eq:Att-L} gives
\eqref{eq:HelloC^i}, another combination appears more novel:\begin{equation}
\overline{\partial}\mathcal{G}=\frac{1}{2}\left[L^{I}\partial_{I}\Omega_{3}-\overline{C}\overline{\Omega}_{3}\right].\label{eq:ExactG}\end{equation}
The operator introduced in \eqref{eq:DiffOp} is nilpotent,\begin{equation}
\int\partial\wedge\partial=\frac{\partial}{\partial e_{I}}\frac{\partial}{\partial m^{I}}-\frac{\partial}{\partial m^{I}}\frac{\partial}{\partial e_{I}}=0\,,\end{equation}
so we find that\begin{equation}
\int\overline{\partial}\wedge\left[L^{I}\partial_{I}\Omega_{3}-\overline{C}\overline{\Omega}_{3}\right]=0\,,\end{equation}
in other words $L^{I}\partial_{I}\Omega_{3}-\overline{C}\overline{\Omega}_{3}$
is $\overline{\partial}-$closed. Indeed, according to \eqref{eq:ExactG}
it is $\overline{\partial}-$exact. This observation may motivate
the introduction of the generating function $\mathcal{G}$ even in
cases where the $F_{I}$ are not globally well-defined.

\section{\label{sec:STUSolution}An Explicit Solution of the Attractor Equations}

In this section we find an explicit solution to the attractor equations
for a particular prepotential:\begin{equation}
F=\frac{Z^{1}Z^{2}Z^{3}}{Z^{0}}\,.\label{eq:STU-prepotential}\end{equation}
This prepotential appears frequently in the supergravity literature
as the STU model \citep{Duff:1995sm,Behrndt:1996hu,LopesCardoso:1996yk,Behrndt:1996jn}, while in the
flux compactification literature it appears as the untwisted sector
of a $T^{6}/\mathbb{Z}^{2}\times\mathbb{Z}^{2}\approx T^{2}\times T^{2}\times T^{2}$
orbifold \citep{Lust:2004fi,Blumenhagen:2006ci}. Because it is a
truncation of $\mathcal{N}=8$ supergravity it has a number of useful
symmetries. On the other hand, it shares many features with more generic
prepotentials, and so is of broader interest than the pure $\mathcal{N}=8$
model.

We first write down the attractor equations explicitly for an arbitrary
set of fluxes. For a \emph{subset} of all possible fluxes, we are
able to solve the attractor equations, finding explicit expressions
for the complex structure moduli and $\tau.$ We then compute the
generating function $\mathcal{G}$ and the gravitino mass, and verify
that the proposed relationship between them \eqref{eq:GeneralG} holds
in this case. We conclude with a discussion of the U-duality group
for this model, and describe how to generalize the solution for our
subset of fluxes to a solution for general fluxes.

\subsection{Symplectic Section and Electric Attractor Equations}

In order to make the attractor equations \eqref{eq:dAtt-1}-\eqref{eq:3-0-Constraint}
completely explicit, we need to specify the symplectic section $\left\{ Z^{I},F_{I}\right\} .$
In the present case the $F_{I}$ are just derivatives of the prepotential
\eqref{eq:STU-prepotential}:\begin{eqnarray}
F_{I} & = & \frac{\partial F}{\partial Z^{I}}\,,\label{eq:FI-STU}\end{eqnarray}
with $I=0,1,2,3.$ We substitute \eqref{eq:FI-STU} into \eqref{eq:V-def}
to find the generating function in the mixed ensemble:\begin{eqnarray}
\mathcal{V}\left(m^{I},\varphi^{I},\tau\right) & = & 2\mbox{Im}\left(\tau\right)C\left[-L^{0}\frac{Z^{1}Z^{2}Z^{3}}{\left(Z^{0}\right)^{2}}+L^{1}\frac{Z^{2}Z^{3}}{Z^{0}}+L^{2}\frac{Z^{3}Z^{1}}{Z^{0}}+L^{3}\frac{Z^{1}Z^{2}}{Z^{0}}\right]\,.\\
 & = & \frac{1}{2\mbox{Im}\left(\tau\right)\left(\overline{m}^{0}+\overline{\varphi}^{0}\right)}\left\{ \frac{-m^{0}+\varphi^{0}}{\overline{m}^{0}+\overline{\varphi}^{0}}\left[\overline{m}^{1}+\overline{\varphi}^{1}\right]\left[\overline{m}^{2}+\overline{\varphi}^{2}\right]\left[\overline{m}^{3}+\overline{\varphi}^{3}\right]\right.\nonumber \\
 &  & -\left[-m^{1}+\varphi^{1}\right]\left[\overline{m}^{2}+\overline{\varphi}^{2}\right]\left[\overline{m}^{3}+\overline{\varphi}^{3}\right]-\left[\overline{m}^{1}+\overline{\varphi}^{1}\right]\left[-m^{2}+\varphi^{2}\right]\left[\overline{m}^{3}+\overline{\varphi}^{3}\right]\nonumber \\
 &  & \left.-\left[\overline{m}^{1}+\overline{\varphi}^{1}\right]\left[\overline{m}^{2}+\overline{\varphi}^{2}\right]\left[-m^{3}+\varphi^{3}\right]\vphantom{\frac{m^{0}}{\overline{m}^{0}}}\right\} .\end{eqnarray}
Since $\mathcal{V}$ is a function of magnetic charges and electric
potentials, we substituted in \eqref{eq:CZI-complex} and \eqref{eq:LI-complex}
for the $Z^{I}$ and $L^{I}$. The electric attractor equations \eqref{eq:eih-tau}
and \eqref{eq:eif-tau} require that we differentiate%
\footnote{We could also have substituted our $F_{I}$ directly into the electric
attractor equations \eqref{eq:dAtt-3} and \eqref{eq:dAtt-4}, then
made the change of variables \eqref{eq:Z-expr} and \eqref{eq:L-expr}.
This gives an identical result, indicating that our $\mbox{Im}\left(\mathcal{V}\right)$
correctly generates the electric attractor equations.%
} $\mbox{Im}\left(\mathcal{V}\right)$: \begin{eqnarray}
\overline{e}_{0} & = & -2i\mbox{Im}\left(\tau\right)\frac{\partial}{\partial\varphi^{0}}\frac{\mathcal{V}-\overline{\mathcal{V}}}{2i}\\
 & = & -\frac{1}{2\left(\overline{m}^{0}+\overline{\varphi}^{0}\right)^{2}}\left[\overline{m}^{1}+\overline{\varphi}^{1}\right]\left[\overline{m}^{2}+\overline{\varphi}^{2}\right]\left[\overline{m}^{3}+\overline{\varphi}^{3}\right]\nonumber \\
 &  & -\frac{1}{2\left(m^{0}+\varphi^{0}\right)^{2}}\left\{ 2\frac{-\overline{m}^{0}+\overline{\varphi}^{0}}{m^{0}+\varphi^{0}}\left[m^{1}+\varphi^{1}\right]\left[m^{2}+\varphi^{2}\right]\left[m^{3}+\varphi^{3}\right]\right.\nonumber \\
 &  & -\left[-\overline{m}^{1}+\overline{\varphi}^{1}\right]\left[m^{2}+\varphi^{2}\right]\left[m^{3}+\varphi^{3}\right]-\left[m^{1}+\varphi^{1}\right]\left[-\overline{m}^{2}+\overline{\varphi}^{2}\right]\left[m^{3}+\varphi^{3}\right]\nonumber \\
 &  & \left.-\left[m^{1}+\varphi^{1}\right]\left[m^{2}+\varphi^{2}\right]\left[-\overline{m}^{3}+\overline{\varphi}^{3}\right]\vphantom{\frac{\overline{m}^{0}}{m^{0}}}\right\} ,\label{eq:e0-gen}\end{eqnarray}
\begin{eqnarray}
\overline{e}_{1} & = & -2i\mbox{Im}\left(\tau\right)\frac{\partial}{\partial\varphi^{1}}\frac{\mathcal{V}-\overline{\mathcal{V}}}{2i}\\
 & = & \frac{1}{2\left(\overline{m}^{0}+\overline{\varphi}^{0}\right)}\left[\overline{m}^{2}+\overline{\varphi}^{2}\right]\left[\overline{m}^{3}+\overline{\varphi}^{3}\right]\nonumber \\
 &  & +\frac{1}{2\left(m^{0}+\varphi^{0}\right)}\left\{ \frac{-\overline{m}^{0}+\overline{\varphi}^{0}}{m^{0}+\varphi^{0}}\left[m^{2}+\varphi^{2}\right]\left[m^{3}+\varphi^{3}\right]\right.\nonumber \\
 &  & \left.-\left[-\overline{m}^{2}+\overline{\varphi}^{2}\right]\left[m^{3}+\varphi^{3}\right]-\left[m^{2}+\varphi^{2}\right]\left[-\overline{m}^{3}+\overline{\varphi}^{3}\right]\vphantom{\frac{\overline{m}^{0}}{m^{0}}}\right\} ,\label{eq:ei-gen}\end{eqnarray}
where the $\varphi^{I}-$derivatives are defined analogous to $e_{I}-$derivatives
\eqref{eq:e-deriv-Complex}. The equations for $\overline{e}_{2}$
and $\overline{e}_{3}$ are cyclic permutations of \eqref{eq:ei-gen},
so we have a system of four complex equations. 

We also need to make the constraint \eqref{eq:3-0-Constraint} explicit.
For the prepotential \eqref{eq:STU-prepotential}, it reduces to\begin{eqnarray}
0 & = & L^{I}\overline{F}_{I}-L^{I}F_{IJ}\overline{Z}^{J}\\
 & = & -L^{0}\frac{\overline{Z}^{1}\overline{Z}^{2}\overline{Z}^{3}}{\left(\overline{Z}^{0}\right)^{2}}+\left[L^{1}\frac{\overline{Z}^{2}\overline{Z}^{3}}{\overline{Z}^{0}}+\mbox{cyc.}\right]-2L^{0}\frac{Z^{1}Z^{2}Z^{3}}{\left(Z^{0}\right)^{3}}\overline{Z}^{0}+\left[L^{0}\frac{Z^{1}Z^{2}}{\left(Z^{0}\right)^{2}}\overline{Z}^{3}+\mbox{cyc.}\right]\nonumber \\
 &  & +\left[L^{1}\frac{Z^{2}Z^{3}}{\left(Z^{0}\right)^{2}}\overline{Z}^{0}+\mbox{cyc.}\right]-\left[L^{1}\frac{Z^{2}}{Z^{0}}\overline{Z}^{3}+L^{1}\frac{Z^{3}}{Z^{0}}\overline{Z}^{2}+\mbox{cyc.}\right].\end{eqnarray}
After we substitute in \eqref{eq:Z-expr} and \eqref{eq:L-expr} this
expands out to\begin{eqnarray}
0 & = & -\left(-m^{0}+\varphi^{0}\right)\frac{\left(m^{1}+\varphi^{1}\right)\left(m^{2}+\varphi^{2}\right)\left(m^{3}+\varphi^{3}\right)}{\left(m^{0}+\varphi^{0}\right)^{2}}\nonumber \\
 &  & +\left[\left(-m^{1}+\varphi^{1}\right)\frac{\left(m^{2}+\varphi^{2}\right)\left(m^{3}+\varphi^{3}\right)}{m^{0}+\varphi^{0}}+\mbox{cyc.}\right]\nonumber \\
 &  & -2\left(-m^{0}+\varphi^{0}\right)\frac{\left(\overline{m}^{1}+\overline{\varphi}^{1}\right)\left(\overline{m}^{2}+\overline{\varphi}^{2}\right)\left(\overline{m}^{3}+\overline{\varphi}^{3}\right)}{\left(\overline{m}^{0}+\overline{\varphi}^{0}\right)^{3}}\left(m^{0}+\varphi^{0}\right)\nonumber \\
 &  & +\left[\left(-m^{0}+\varphi^{0}\right)\frac{\left(\overline{m}^{1}+\overline{\varphi}^{1}\right)\left(\overline{m}^{2}+\overline{\varphi}^{2}\right)}{\left(\overline{m}^{0}+\overline{\varphi}^{0}\right)^{2}}\left(m^{3}+\varphi^{3}\right)+\mbox{cyc.}\right]\nonumber \\
 &  & +\left[\left(-m^{1}+\varphi^{1}\right)\frac{\left(\overline{m}^{2}+\overline{\varphi}^{2}\right)\left(\overline{m}^{3}+\overline{\varphi}^{3}\right)}{\left(\overline{m}^{0}+\overline{\varphi}^{0}\right)^{2}}\left(m^{0}+\varphi^{0}\right)+\mbox{cyc.}\right]\label{eq:constraint-gen}\\
 &  & -\left[\left(-m^{1}+\varphi^{1}\right)\frac{\overline{m}^{2}+\overline{\varphi}^{2}}{\overline{m}^{0}+\overline{\varphi}^{0}}\left(m^{3}+\varphi^{3}\right)+\left(-m^{1}+\varphi^{1}\right)\frac{\overline{m}^{3}+\overline{\varphi}^{3}}{\overline{m}^{0}+\overline{\varphi}^{0}}\left(m^{2}+\varphi^{2}\right)+\mbox{cyc.}\right].\nonumber \end{eqnarray}
This appears to be another high-order polynomial equation in many
variables.

We need to invert \eqref{eq:e0-gen}, \eqref{eq:ei-gen}, and \eqref{eq:constraint-gen}
and find both the electric potentials $\varphi^{I}$ and $\tau$ as
functions of the electric and magnetic fluxes. Doing this by brute
force would be quite challenging, as each equation is at least cubic
in the potentials. Although we have written the attractor equations
in terms of complex potentials and fluxes they are clearly not holomorphic
in the potentials, so even counting the number of distinct solutions
(sometimes called {}``area codes'' \citep{Moore:1998pn,Kallosh:1999mz,Denef:2000nb,Denef:2001xn,Giryavets:2005nf})
for general fluxes appears difficult. In the following we will find
a solution to these equations using the ideas developed in section
\eqref{sec:Solving-the-Flux}.

\subsection{\label{sub:FluxReduction}Reduction to Eight Fluxes}

Much of the difficulty in solving \eqref{eq:e0-gen}, \eqref{eq:ei-gen},
and \eqref{eq:constraint-gen} arises from their dependence on both
$m^{I},$ $\varphi^{I},$ \emph{and} $\overline{m}^{I},$ $\overline{\varphi}^{I}.$
Things simplify quite a bit if we set $m_{h}^{0}=m_{f}^{i}=e_{0}^{f}=e_{i}^{h}=0,$
and make the \emph{ansatz} that $\mbox{Re}\left(\tau\right)=\phi_{h}^{0}=\phi_{f}^{I}=0,$
so that the complex fluxes and potentials become:\begin{eqnarray}
m^{0} & = & m_{f}^{0}\,,\\
m^{i} & = & -i\mbox{Im}\left(\tau\right)m_{h}^{i}\,,\\
e_{0} & = & -i\mbox{Im}\left(\tau\right)e_{0}^{h}\,,\\
e_{i} & = & e_{i}^{f}\,,\\
\varphi^{0} & = & \phi_{f}^{0}\,,\\
\varphi^{i} & = & -i\mbox{Im}\left(\tau\right)\phi_{h}^{i}\,.\end{eqnarray}
This makes it easy to take the complex conjugate of a flux or potential:
$\overline{m}^{0}=m^{0},$ $\overline{e}_{i}=e_{i},$ $\overline{\varphi}^{0}=\varphi^{0},$
$\overline{m}^{i}=-m^{i},$ $\overline{e}_{0}=-e_{0},$ and $\overline{\varphi}^{i}=-\varphi^{i}.$

If we apply these restrictions to \eqref{eq:e0-gen}, \eqref{eq:ei-gen},
and \eqref{eq:constraint-gen} we find:\begin{eqnarray}
e_{0} & = & -\frac{\left(m^{1}+\varphi^{1}\right)\left(m^{2}+\varphi^{2}\right)\left(m^{3}+\varphi^{3}\right)}{2\left(m^{0}+\varphi^{0}\right)^{2}}\left\{ 1-2\frac{-m^{0}+\varphi^{0}}{m^{0}+\varphi^{0}}-\frac{-m^{1}+\varphi^{1}}{m^{1}+\varphi^{1}}\right.\nonumber \\
 &  & \left.-\frac{-m^{2}+\varphi^{2}}{m^{2}+\varphi^{2}}-\frac{-m^{3}+\varphi^{3}}{m^{3}+\varphi^{3}}\right\} ,\label{eq:eo-reduced}\\
e_{1} & = & \frac{\left(m^{2}+\varphi^{2}\right)\left(m^{3}+\varphi^{3}\right)}{2\left(m^{0}+\varphi^{0}\right)}\left\{ 1+\frac{-m^{0}+\varphi^{0}}{m^{0}+\varphi^{0}}+\frac{-m^{2}+\varphi^{2}}{m^{2}+\varphi^{2}}+\frac{-m^{3}+\varphi^{3}}{m^{3}+\varphi^{3}}\right\} ,\label{eq:e-reduced}\\
0 & = & \frac{\left(m^{1}+\varphi^{1}\right)\left(m^{2}+\varphi^{2}\right)\left(m^{3}+\varphi^{3}\right)}{2\left(m^{0}+\varphi^{0}\right)^{2}}\left[\frac{-m^{0}+\varphi^{0}}{m^{0}+\varphi^{0}}+\frac{-m^{1}+\varphi^{1}}{m^{1}+\varphi^{1}}\right.\nonumber \\
 &  & \left.+\frac{-m^{2}+\varphi^{2}}{m^{2}+\varphi^{2}}+\frac{-m^{3}+\varphi^{3}}{m^{3}+\varphi^{3}}\right].\label{eq:constraint-reduced}\end{eqnarray}
Note that the same prefactor appears in \eqref{eq:eo-reduced} and
\eqref{eq:constraint-reduced}. So long as $e_{0}\neq0,$ we conclude
that the factor in square brackets in \eqref{eq:constraint-reduced}
must vanish. We can apply this to \eqref{eq:eo-reduced} and \eqref{eq:e-reduced}
to arrive at a simpler set of equations:\begin{eqnarray}
e_{0} & = & -\frac{\left(m^{1}+\varphi^{1}\right)\left(m^{2}+\varphi^{2}\right)\left(m^{3}+\varphi^{3}\right)}{\left(m^{0}+\varphi^{0}\right)^{3}}m^{0}\,,\label{eq:e0-solve}\\
e_{1} & = & \frac{\left(m^{2}+\varphi^{2}\right)\left(m^{3}+\varphi^{3}\right)}{\left(m^{0}+\varphi^{0}\right)\left(m^{1}+\varphi^{1}\right)}m^{1}\,,\label{eq:e-solve}\\
0 & = & \frac{-m^{0}+\varphi^{0}}{m^{0}+\varphi^{0}}+\frac{-m^{1}+\varphi^{1}}{m^{1}+\varphi^{1}}+\frac{-m^{2}+\varphi^{2}}{m^{2}+\varphi^{2}}+\frac{-m^{3}+\varphi^{3}}{m^{3}+\varphi^{3}}\,.\label{eq:constraint-solve}\end{eqnarray}
As usual, expressions for $e_{2}$ and $e_{3}$ arise from cyclic
permutations of \eqref{eq:e-solve}. In the next section we will explicitly
invert these equations.

\subsection{\label{sub:Moduli-Reduced}Moduli, Potentials, and Mass Parameters
(Reduced Fluxes)}

We begin by solving for the physical complex structure moduli,\begin{equation}
z^{i}\equiv\frac{Z^{i}}{Z^{0}}=\frac{\overline{m}^{i}+\overline{\varphi}^{i}}{\overline{m}^{0}+\overline{\varphi}^{0}}=-\frac{m^{i}+\varphi^{i}}{m^{0}+\varphi^{0}}\,.\label{eq:first-zs}\end{equation}
The ratio of \eqref{eq:e0-solve} and \eqref{eq:e-solve} can be solved
for the $z^{i}:$\begin{eqnarray}
\frac{e_{i}}{e_{0}} & = & -\left(\frac{m^{0}+\varphi^{0}}{m^{i}+\varphi^{i}}\right)^{2}\frac{m^{i}}{m^{0}}=-\frac{1}{\left(z^{i}\right)^{2}}\frac{m^{i}}{m^{0}}\,.\end{eqnarray}
In order to avoid awkward branch cuts when we take the square root,
we will carefully analyze the signs on the charges. If we insert the
real charges and potentials into the previous expression,\begin{equation}
\frac{e_{i}^{f}}{e_{0}^{h}}=-\frac{m_{h}^{i}}{m_{f}^{0}}\left(\frac{m_{f}^{0}+\phi_{f}^{0}}{m_{h}^{i}+\phi_{h}^{i}}\right)^{2},\end{equation}
 we find that $e_{i}^{f}m_{f}^{0}/e_{0}^{h}m_{h}^{i}<0,$ and thus
that $e_{i}m^{0}/e_{0}m^{i}>0.$ We must also consider the Kähler
potential \eqref{eq:K-z} with the prepotential \eqref{eq:STU-prepotential}.
Evaluating it, we find\begin{eqnarray}
K_{z} & = & -\log\left|Z^{0}\right|^{2}-\log\left[-8\mbox{Im}\left(z^{1}\right)\mbox{Im}\left(z^{2}\right)\mbox{Im}\left(z^{3}\right)\right].\label{eq:Kz-STU}\end{eqnarray}
The condition that the volume of each of the underlying $T^{2}$'s
is positive requires $\mbox{Im}\left(z^{i}\right)<0,$ which in turn
implies that $K_{z}$ is real. This determines the expression for
the modulus:\begin{equation}
z^{i}=-i\sqrt{\frac{e_{0}m^{i}}{m^{0}e_{i}}}=-i\mbox{Im}\left(\tau\right)\sqrt{-\frac{e_{0}^{h}m_{h}^{i}}{m_{f}^{0}e_{i}^{f}}}\,.\label{eq:early-z}\end{equation}
In order to make this completely explicit we must solve for $\mbox{Im}\left(\tau\right),$
so we will do that next.

We can use \eqref{eq:early-z} to simplify \eqref{eq:e0-solve}: \begin{equation}
e_{0}=z^{1}z^{2}z^{3}m^{0}=i\sqrt{\frac{\left(e_{0}\right)^{3}m^{1}m^{2}m^{3}}{\left(m^{0}\right)^{3}e_{1}e_{2}e_{3}}}m^{0}\,.\end{equation}
All dependence on the potentials has been eliminated, so this is a
single equation that determines $\mbox{Im}\left(\tau\right).$ Substituting
in real quantities, we find \begin{eqnarray}
1 & = & -\mbox{sgn}\left(m_{f}^{0}e_{0}^{h}\right)\sqrt{-\mbox{Im}\left(\tau\right)^{4}\frac{e_{0}^{h}m_{h}^{1}m_{h}^{2}m_{h}^{3}}{m_{f}^{0}e_{1}^{f}e_{2}^{f}e_{3}^{f}}}\,.\label{eq:imtau-bonus}\end{eqnarray}
Note that the $\mbox{sgn}\left(m_{f}^{0}e_{0}^{h}\right)$ appeared
when we pulled the factor of $m_{f}^{0}/e_{0}^{h}$ under the square
root. We now find that \begin{eqnarray}
\mbox{Im}\left(\tau\right) & = & \left(-\frac{m_{f}^{0}e_{1}^{f}e_{2}^{f}e_{3}^{f}}{e_{0}^{h}m_{h}^{1}m_{h}^{2}m_{h}^{3}}\right)^{1/4},\label{eq:imtau-reduced}\end{eqnarray}
where the physical condition $\mbox{Im}\left(\tau\right)=e^{-\phi}$
dictates that we use the real, positive branch, and implies that $K_{\tau}$
\eqref{eq:KahlerPot} is real%
\footnote{It is somewhat awkward that our Kähler potential requires $\mbox{Im}\left(\tau\right)>0$
but $\mbox{Im}\left(z^{i}\right)<0,$ especially if we want to consider
this model as a compactification of F-theory. On the other hand, our
conventions are self-consistent, and chosen to agree with the bulk
of the literature on flux compactifications.%
}. 

Equation \eqref{eq:imtau-bonus} also implies that $\mbox{sgn}\left(m_{f}^{0}e_{0}^{h}\right)=-1.$
We can combine this with our earlier result that $\mbox{sgn}\left(m_{f}^{0}e_{0}^{h}m_{h}^{i}e_{i}^{f}\right)=-1$
to find a complete set of sign restrictions:\begin{equation}
-\mbox{sgn}\left(m_{f}^{0}e_{0}^{h}\right)=\mbox{sgn}\left(m_{h}^{1}e_{1}^{f}\right)=\mbox{sgn}\left(m_{h}^{2}e_{2}^{f}\right)=\mbox{sgn}\left(m_{h}^{3}e_{3}^{f}\right)=+1.\label{eq:i-restrict}\end{equation}
Only 1/16 of the possible fluxes satisfy the physical conditions we
have imposed. It is interesting to consider what might happen if we
relaxed these sign restrictions. Suppose we chose signs that violated
some of the conditions in \eqref{eq:i-restrict}, but satisfied the
\emph{product} of those conditions. The Kähler potential \eqref{eq:KahlerPot}
would still be real, so we would still have solutions to the
ISD condition, at least formally. The caveat is that the complex structures of some
of the $T^2$'s would no longer be in the upper half-plane and/or the sign of the string coupling would be negative. At a minimum, then, we would have to
give up the conventional geometrical interpretation of the moduli.  Going even further,
we can consider signs such that the \emph{product} of the conditions
in \eqref{eq:i-restrict} are violated. Then the Kähler potential
\eqref{eq:KahlerPot} would not be real and it is not clear that the
proposed solution would, in fact, be a solution. Indeed, for such flux assignments
there may not be any solutions to the ISD conditions at all. In the
following we will analyze only the clearly physical solutions that
satisfy \eqref{eq:i-restrict}. 

We can compare our restrictions with a more familiar one \citep{Douglas:2006es}.
If we assume that the attractor equations can be satisfied, i.e. \eqref{eq:HelloC^i},
then \begin{eqnarray}
\int F_{3}\wedge H_{3} & = & \frac{1}{2i\mbox{Im}\left(\tau\right)}\int G_{3}\wedge\overline{G}_{3}\\
 & = & \frac{e^{-K_{z}}}{2\mbox{Im}\left(\tau\right)}\left[\left|C\right|^{2}+\left|C^{i}\right|^{2}\right],\end{eqnarray}
and thus $\int F_{3}\wedge H_{3}$ is positive. The sign restrictions
\eqref{eq:i-restrict} are consistent with this, but stronger. If
we evaluate $\int F_{3}\wedge H_{3}$ for our reduced fluxes, \begin{equation}
\int F_{3}\wedge H_{3}=-e_{0}^{h}m_{f}^{0}+e_{i}^{f}m_{h}^{i}\,,\label{eq:FH-reduced}\end{equation}
we see that the sign restrictions require that \emph{each term} be
positive.

Having determined $\mbox{Im}\left(\tau\right)$ and the sign restrictions
on the various fluxes, \eqref{eq:early-z} gives explicit expressions
for the complex structure moduli:\begin{equation}
z^{1}=-i\left[\left(-\frac{e_{0}^{h}}{m_{f}^{0}}\right)\left(\frac{m_{h}^{1}}{e_{1}^{f}}\right)\left(\frac{e_{2}^{f}}{m_{h}^{2}}\right)\left(\frac{e_{3}^{f}}{m_{h}^{3}}\right)\right]^{1/4},\label{eq:zi-reduced}\end{equation}
and cyclic permutations. These explicit expressions for the physical
moduli, along with the dilaton \eqref{eq:imtau-reduced} and the restrictions
on the fluxes \eqref{eq:i-restrict}, are some of the principal results
of this example.

Up to this point we have solved for the moduli and derived a set of
restrictions on the fluxes, but we haven't yet solved for the potentials.
The only equation that we haven't solved is the constraint \eqref{eq:constraint-solve},
so let's turn our attention there. We can rewrite that equation as\begin{equation}
m^{0}+\varphi^{0}=\frac{m^{0}}{2}\left\{ 1+\frac{m^{1}}{m^{0}}\frac{m^{0}+\varphi^{0}}{m^{1}+\varphi^{1}}+\frac{m^{2}}{m^{0}}\frac{m^{0}+\varphi^{0}}{m^{2}+\varphi^{2}}+\frac{m^{3}}{m^{0}}\frac{m^{0}+\varphi^{0}}{m^{3}+\varphi^{3}}\right\} .\end{equation}
Combining \eqref{eq:first-zs} and \eqref{eq:early-z}, we find\begin{equation}
m^{0}+\varphi^{0}=\frac{m^{0}}{2}\left\{ 1-i\frac{m^{1}}{m^{0}}\sqrt{\frac{m^{0}e_{1}}{e_{0}m^{1}}}-i\frac{m^{2}}{m^{0}}\sqrt{\frac{m^{0}e_{2}}{e_{0}m^{2}}}-i\frac{m^{3}}{m^{0}}\sqrt{\frac{m^{0}e_{3}}{e_{0}m^{3}}}\right\} .\end{equation}
We now rewrite this in terms of real quantities:\begin{eqnarray}
\phi_{f}^{0} & = & \frac{m_{f}^{0}}{2}\left\{ -1-\mbox{sgn}\left(m_{f}^{0}m_{h}^{1}\right)\sqrt{-\frac{m_{h}^{1}e_{1}^{f}}{e_{0}^{h}m_{f}^{0}}}-\mbox{sgn}\left(m_{f}^{0}m_{h}^{2}\right)\sqrt{-\frac{m_{h}^{2}e_{2}^{f}}{e_{0}^{h}m_{f}^{0}}}\right.\nonumber \\
 &  & \left.-\mbox{sgn}\left(m_{f}^{0}m_{h}^{3}\right)\sqrt{-\frac{m_{h}^{3}e_{3}^{f}}{e_{0}^{h}m_{f}^{0}}}\right\} .\label{eq:phi0-solution}\end{eqnarray}
If we again use the relation between $m^{0}+\varphi^{0}$ and $m^{1}+\varphi^{1},$
\eqref{eq:first-zs}, we find the following expression for $\phi_{h}^{1}:$\begin{eqnarray}
\phi_{h}^{1} & = & \frac{m_{h}^{1}}{2}\left\{ -1-\mbox{sgn}\left(m_{h}^{1}m_{f}^{0}\right)\sqrt{-\frac{m_{f}^{0}e_{0}^{h}}{m_{h}^{1}e_{1}^{f}}}+\mbox{sgn}\left(m_{h}^{1}m_{h}^{2}\right)\sqrt{\frac{m_{h}^{2}e_{2}^{f}}{m_{h}^{1}e_{1}^{f}}}\right.\nonumber \\
 &  & \left.+\mbox{sgn}\left(m_{h}^{1}m_{h}^{3}\right)\sqrt{\frac{m_{h}^{3}e_{3}^{f}}{m_{h}^{1}e_{1}^{f}}}\right\} .\label{eq:phii-solution}\end{eqnarray}
This completes our inversion of \eqref{eq:e0-solve}, \eqref{eq:e-solve},
and \eqref{eq:constraint-solve}. 

We emphasized earlier in this paper that the attractor equations include
the mass parameters $C^{i}$ on equal terms with the moduli $z^{i}.$
With \eqref{eq:phi0-solution} and \eqref{eq:phii-solution} in hand,
it is straightforward to compute the $C^{i}.$ We first insert our
$z^{i}=Z^{i}/Z^{0}$ into \eqref{eq:CtoL} to make the relationship
between the $C^{i}$ and $L^{I}$ explicit:\begin{equation}
C^{i}Z^{0}=-z^{i}L^{0}+L^{i}\,.\label{eq:LtoC-explicit}\end{equation}
Note that the combination $C^{i}Z^{0}$ is Kähler-invariant, while
$C^{i}$ alone is not. If we substitute \eqref{eq:L-expr} and \eqref{eq:early-z}
into \eqref{eq:LtoC-explicit}, we find\begin{eqnarray}
C^{1}Z^{0} & = & \sqrt{-\frac{e_{0}^{h}m_{h}^{1}}{m_{f}^{0}e_{1}^{f}}}\frac{1}{2}\left(-m_{f}^{0}+\phi_{f}^{0}\right)-\frac{1}{2}\left(-m_{h}^{1}+\phi_{h}^{1}\right)\\
 & = & \frac{1}{2}\left[\mbox{sgn}\left(m_{f}^{0}m_{h}^{1}\right)\frac{m_{h}^{1}}{m_{f}^{0}}\sqrt{-\frac{m_{f}^{0}e_{0}^{h}}{m_{h}^{1}e_{1}^{f}}}\left(-m_{f}^{0}+\phi_{f}^{0}\right)-\left(-m_{h}^{1}+\phi_{h}^{1}\right)\right]\\
 & = & \frac{m_{h}^{1}}{4}\mbox{sgn}\left(m_{f}^{0}m_{h}^{1}\right)\sqrt{-\frac{m_{f}^{0}e_{0}^{h}}{m_{h}^{1}e_{1}^{f}}}\left[-3-\sum_{i=1}^{3}\mbox{sgn}\left(m_{f}^{0}m_{h}^{i}\right)\sqrt{-\frac{m_{h}^{i}e_{i}^{f}}{e_{0}^{h}m_{f}^{0}}}\right]\nonumber \\
 &  & -\frac{m_{h}^{1}}{4}\left[-3-\mbox{sgn}\left(m_{h}^{1}m_{f}^{0}\right)\sqrt{-\frac{m_{f}^{0}e_{0}^{h}}{m_{h}^{1}e_{1}^{f}}}+\mbox{sgn}\left(m_{h}^{1}m_{h}^{2}\right)\sqrt{\frac{m_{h}^{2}e_{2}^{f}}{m_{h}^{1}e_{1}^{f}}}+\mbox{sgn}\left(m_{h}^{1}m_{h}^{3}\right)\sqrt{\frac{m_{h}^{3}e_{3}^{f}}{m_{h}^{1}e_{1}^{f}}}\right]\nonumber \\
 & = & \frac{m_{h}^{1}}{4}\left[-3\mbox{sgn}\left(m_{f}^{0}m_{h}^{1}\right)\sqrt{-\frac{m_{f}^{0}e_{0}^{h}}{m_{h}^{1}e_{1}^{f}}}-1-\mbox{sgn}\left(m_{h}^{1}m_{h}^{2}\right)\sqrt{\frac{m_{h}^{2}e_{2}^{f}}{m_{h}^{1}e_{1}^{f}}}-\mbox{sgn}\left(m_{h}^{1}m_{h}^{3}\right)\sqrt{\frac{m_{h}^{3}e_{3}^{f}}{m_{h}^{1}e_{1}^{f}}}\right]\nonumber \\
 &  & -\frac{m_{h}^{1}}{4}\left[-3-\mbox{sgn}\left(m_{h}^{1}m_{f}^{0}\right)\sqrt{-\frac{m_{f}^{0}e_{0}^{h}}{m_{h}^{1}e_{1}^{f}}}+\mbox{sgn}\left(m_{h}^{1}m_{h}^{2}\right)\sqrt{\frac{m_{h}^{2}e_{2}^{f}}{m_{h}^{1}e_{1}^{f}}}+\mbox{sgn}\left(m_{h}^{1}m_{h}^{3}\right)\sqrt{\frac{m_{h}^{3}e_{3}^{f}}{m_{h}^{1}e_{1}^{f}}}\right]\nonumber \\
 & = & \frac{m_{h}^{1}}{2}\left[1-\mbox{sgn}\left(m_{f}^{0}m_{h}^{1}\right)\sqrt{-\frac{m_{f}^{0}e_{0}^{h}}{m_{h}^{1}e_{1}^{f}}}-\mbox{sgn}\left(m_{h}^{1}m_{h}^{2}\right)\sqrt{\frac{m_{h}^{2}e_{2}^{f}}{m_{h}^{1}e_{1}^{f}}}-\mbox{sgn}\left(m_{h}^{1}m_{h}^{3}\right)\sqrt{\frac{m_{h}^{3}e_{3}^{f}}{m_{h}^{1}e_{1}^{f}}}\right].\label{eq:ci-explicit}\end{eqnarray}
If one wishes to compute the fermion and scalar mass matrices explicitly,
these expressions can be substituted into \eqref{eq:MassMatrix},
\eqref{eq:Scalar1}, and \eqref{eq:Scalar2}.

\subsection{\label{sub:GenFunctions-Reduced}Generating Functions (Reduced Fluxes)}

One of the principal results of this paper is that the attractor behavior
of these flux compactifications is governed by a single function $\mathcal{G}.$
In this section we compute this function for our reduced fluxes. We
will then verify the simple relationship between $\mathcal{G}$ and
the gravitino mass. 

We begin with $\mbox{Im}\left(\mathcal{V}\right).$ If we substitute
our $F_{I}$ into \eqref{eq:V-def}, we find

\begin{eqnarray}
\mbox{Im}\left(\mathcal{V}\right) & = & 2\mbox{Im}\left(\tau\right)\mbox{Im}\left\{ C\frac{Z^{1}Z^{2}Z^{3}}{Z^{0}}\left[-\frac{L^{0}}{Z^{0}}+\frac{L^{1}}{Z^{1}}+\frac{L^{2}}{Z^{2}}+\frac{L^{3}}{Z^{3}}\right]\right\} \\
 & = & 2\mbox{Im}\left(\tau\right)\mbox{Im}\left\{ -C^{2}\frac{Z^{1}Z^{2}Z^{3}}{Z^{0}}\left[\frac{-m^{0}+\varphi^{0}}{m^{0}+\varphi^{0}}+\frac{-m^{1}+\varphi^{1}}{m^{1}+\varphi^{1}}\right.\right.\nonumber \\
 &  & \left.\left.+\frac{-m^{2}+\varphi^{2}}{m^{2}+\varphi^{2}}+\frac{-m^{3}+\varphi^{3}}{m^{3}+\varphi^{3}}\right]\right\} .\end{eqnarray}
The term in square brackets is just the constraint \textbf{\eqref{eq:constraint-solve}}
so $\mbox{Im}\left(\mathcal{V}\right)=0.$ If we substitute this into
\eqref{eq:G-def}, we  find for our reduced fluxes \begin{equation}
\mathcal{G}=e_{0}^{h}\phi_{f}^{0}-e_{i}^{f}\phi_{h}^{i}\,.\end{equation}
We compute each term separately:\begin{eqnarray}
e_{0}^{h}\phi_{f}^{0} & = & \frac{1}{2}\left\{ -e_{0}^{h}m_{f}^{0}+\mbox{sgn}\left(m_{f}^{0}m_{h}^{1}\right)\sqrt{-e_{0}^{h}m_{f}^{0}}\sqrt{m_{h}^{1}e_{1}^{f}}\right.\nonumber \\
 &  & \left.+\mbox{sgn}\left(m_{f}^{0}m_{h}^{2}\right)\sqrt{-e_{0}^{h}m_{f}^{0}}\sqrt{m_{h}^{2}e_{2}^{f}}+\mbox{sgn}\left(m_{f}^{0}m_{h}^{3}\right)\sqrt{-e_{0}^{h}m_{f}^{0}}\sqrt{m_{h}^{3}e_{3}^{f}}\right\} ,\\
e_{1}^{f}\phi_{h}^{1} & = & \frac{1}{2}\left\{ -e_{1}^{f}m_{h}^{1}-\mbox{sgn}\left(m_{f}^{0}m_{h}^{1}\right)\sqrt{-e_{0}^{h}m_{f}^{0}}\sqrt{m_{h}^{1}e_{1}^{f}}\right.\\
 &  & \left.+\mbox{sgn}\left(m_{h}^{1}m_{h}^{2}\right)\sqrt{m_{h}^{1}e_{1}^{f}}\sqrt{m_{h}^{2}e_{2}^{f}}+\mbox{sgn}\left(m_{h}^{1}m_{h}^{3}\right)\sqrt{m_{h}^{1}e_{1}^{f}}\sqrt{m_{h}^{3}e_{3}^{f}}\right\} .\end{eqnarray}
Putting this together yields \begin{eqnarray}
\mathcal{G} & = & \frac{1}{2}\left[-e_{0}^{h}m_{f}^{0}+e_{i}^{f}m_{h}^{i}\right]+\mbox{sgn}\left(m_{f}^{0}m_{h}^{1}\right)\sqrt{-e_{0}^{h}m_{f}^{0}}\sqrt{m_{h}^{1}e_{1}^{f}}+\mbox{sgn}\left(m_{f}^{0}m_{h}^{2}\right)\sqrt{-e_{0}^{h}m_{f}^{0}}\sqrt{m_{h}^{2}e_{2}^{f}}\nonumber \\
 &  & +\mbox{sgn}\left(m_{f}^{0}m_{h}^{3}\right)\sqrt{-e_{0}^{h}m_{f}^{0}}\sqrt{m_{h}^{3}e_{3}^{f}}-\mbox{sgn}\left(m_{h}^{1}m_{h}^{2}\right)\sqrt{m_{h}^{1}e_{1}^{f}}\sqrt{m_{h}^{2}e_{2}^{f}}\nonumber \\
 &  & -\mbox{sgn}\left(m_{h}^{1}m_{h}^{3}\right)\sqrt{m_{h}^{1}e_{1}^{f}}\sqrt{m_{h}^{3}e_{3}^{f}}-\mbox{sgn}\left(m_{h}^{2}m_{h}^{3}\right)\sqrt{m_{h}^{2}e_{2}^{f}}\sqrt{m_{h}^{3}e_{3}^{f}}\,.\label{eq:G-solution}\end{eqnarray}
The term in square brackets is just $\int F_{3}\wedge H_{3}$ \eqref{eq:FH-reduced},
while the remainder is less familiar. It is precisely what is required
so that $\partial\mathcal{G}/\partial e_{0}^{h}=\phi_{f}^{0}$ and
$\partial\mathcal{G}/\partial e_{i}^{f}=-\phi_{h}^{i},$ as can be
readily verified. It is also closely related to the gravitino mass,
as we will now see.

In order to compute the gravitino mass we substitute \eqref{eq:Kz-STU},
\eqref{eq:imtau-reduced}, \eqref{eq:zi-reduced}, and \eqref{eq:phi0-solution}
into \eqref{eq:FaveGravitino} and simplify

\begin{eqnarray}
\mbox{Vol}^{2}m_{3/2}^{2} & = & -\frac{8\mbox{Im}\left(\tau\right)\mbox{Im}\left(z^{1}\right)\mbox{Im}\left(z^{2}\right)\mbox{Im}\left(z^{3}\right)}{2}\left(\frac{1}{2\mbox{Im}\left(\tau\right)}\right)^{2}\left(m_{f}^{0}+\phi_{f}^{0}\right)^{2}\\
 & = & -\frac{e_{0}^{h}m_{f}^{0}}{4}\left\{ 1-\mbox{sgn}\left(m_{f}^{0}m_{h}^{1}\right)\sqrt{-\frac{m_{h}^{1}e_{1}^{f}}{e_{0}^{h}m_{f}^{0}}}-\mbox{sgn}\left(m_{f}^{0}m_{h}^{2}\right)\sqrt{-\frac{m_{h}^{2}e_{2}^{f}}{e_{0}^{h}m_{f}^{0}}}\right.\nonumber \\
 &  & \left.-\mbox{sgn}\left(m_{f}^{0}m_{h}^{3}\right)\sqrt{-\frac{m_{h}^{3}e_{3}^{f}}{e_{0}^{h}m_{f}^{0}}}\right\} ^{2}\\
 & = & \frac{1}{2}\left\{ \frac{1}{2}\left[-e_{0}^{h}m_{f}^{0}+e_{i}^{f}m_{h}^{i}\right]-\mbox{sgn}\left(m_{f}^{0}m_{h}^{1}\right)\sqrt{-e_{0}^{h}m_{f}^{0}}\sqrt{m_{h}^{1}e_{1}^{f}}\right.\nonumber \\
 &  & -\mbox{sgn}\left(m_{f}^{0}m_{h}^{2}\right)\sqrt{-e_{0}^{h}m_{f}^{0}}\sqrt{m_{h}^{2}e_{2}^{f}}-\mbox{sgn}\left(m_{f}^{0}m_{h}^{3}\right)\sqrt{-e_{0}^{h}m_{f}^{0}}\sqrt{m_{h}^{3}e_{3}^{f}}\nonumber \\
 &  & +\mbox{sgn}\left(m_{h}^{1}m_{h}^{2}\right)\sqrt{m_{h}^{1}e_{1}^{f}}\sqrt{m_{h}^{2}e_{2}^{f}}+\mbox{sgn}\left(m_{h}^{1}m_{h}^{3}\right)\sqrt{m_{h}^{1}e_{1}^{f}}\sqrt{m_{h}^{3}e_{3}^{f}}\nonumber \\
 &  & \left.+\mbox{sgn}\left(m_{h}^{2}m_{h}^{3}\right)\sqrt{m_{h}^{2}e_{2}^{f}}\sqrt{m_{h}^{3}e_{3}^{f}}\right\} \,,\end{eqnarray}
If we compare this with our expression for $\mathcal{G}$ \eqref{eq:G-solution},
we see that they are related by\begin{equation}
\mathcal{G}=\int F_{3}\wedge H_{3}-2\mbox{Vol}^{2}m_{3/2}^{2}\,,\label{eq:GenFromGravitino}\end{equation}
in accord with the general relationship \eqref{eq:GeneralG}.

\subsection{\label{sub:U-Invariants}$U-$Invariants for $F=Z^{1}Z^{2}Z^{3}/Z^{0}$}

The model we are considering enjoys a large set of duality symmetries.
We have not made explicit use of these dualities so far, but in this
section we will show how they may be used to generalize our solution
with only eight fluxes to a solution for the full set of sixteen fluxes.
We take inspiration here from the STU black hole, where consideration
of duality-invariant combinations of the black hole charges led to
a simple expression for the generating function of the potentials
\citep{Behrndt:1996hu,Kallosh:1996uy}. 

One part of the duality group is easily identified if we think of
our prepotential as arising from compactification on $T^{2}\times T^{2}\times T^{2}.$
We can interpret each $z^{i}$ as the modular parameter of the $i$th
torus, and consider modular transformations on each torus. Since the
tori and their associated modular transformations factorize, their
contribution to the U-duality group is just $SL\left(2\right)^{3}.$
This is the symmetry group of the STU black hole \citep{Behrndt:1996hu},
whose charges transform%
\footnote{For details of the action of $SL\left(2\right)^{3}$ on the charges,
see e.g. \citep{Gimon:2007mh}.%
} in the $\left(2,2,2\right)$ of $SL\left(2\right)^{3}.$ 

IIB theories also enjoy an $SL\left(2\right)$ S-duality, independent
of the $SL\left(2\right)^{3}$ that we have already discussed. This
does not factor into discussions of the STU black hole in the IIB
picture%
\footnote{One can also discuss this entirely in the language of $\mathcal{N}=2$
supergravity. In the STU black hole all of the hypermultiplets, including
the universal hypermultiplet, decouple from the attractor flow. On
the other hand the axio-dilaton, which descends from the universal
hypermultiplet, does \emph{not} decouple from the flux attractor.%
}, as the D3-branes that one uses to construct the black hole (see
section \eqref{sub:Black-Hole}) are invariant under S-duality. The
fluxes $H_{3}$ and $F_{3},$ however, transform under S-duality,
so we must consider the larger duality group $SL\left(2\right)^{4},$
under which our fluxes transform as $\left(2,2,2,2\right).$

The discussion of STU black holes in terms of $SL\left(2\right)^{3}$
invariants is relatively straightforward because there is a single
$SL\left(2\right)^{3}$-invariant that one can construct from the
charges \citep{Kallosh:1996uy}. This essentially determines the black
hole entropy, which in turn is the generating function for the electric
and magnetic potentials. On the other hand, one can construct \emph{four}
invariants%
\footnote{More precisely, one can construct exactly four invariants from the
$\left(2,2,2,2\right)$ of $SL\left(2,\mathbb{C}\right)^{4}.$ These
are also invariants of $SL\left(2,\mathbb{R}\right)^{4}$ but additional
invariants might arise when we restrict to the subgroup. Possible
examples include $\mbox{sgn}\left(m_{f}^{0}m_{h}^{i}\right).$ We
also expect some number of discrete invariants to appear upon further
restriction to $SL\left(2,\mathbb{Z}\right)^{4}.$%
} from the $\left(2,2,2,2\right)$ of $SL\left(2\right)^{4}$ \citep{quant-ph/0212069}.
The quadratic $I_{2}=\int F_{3}\wedge H_{3}$ appears in most studies
of IIB flux compactifications, while the other three are less familiar.
Considered as polynomials in the fluxes, there are also two quartics,
$I_{4}^{\left(1\right)}$ and $I_{4}^{\left(2\right)},$ and a sextic,
$I_{6}.$

In section \ref{sub:FluxReduction} we chose a reduced set of fluxes
that allowed us to explicitly solve the attractor equations. One of
our motivations in choosing these particular fluxes was to choose
a combination that left all four $SL\left(2\right)^{4}$ invariants
non-zero and independent. While the general expressions for these
invariants are quite complicated (see \citep{quant-ph/0212069} for
details), they simplify considerably for our reduced fluxes:\begin{eqnarray}
I_{2} & = & \int F_{3}\wedge H_{3}=\left(-m_{f}^{0}e_{0}^{h}\right)+\left(e_{1}^{f}m_{h}^{1}\right)+\left(e_{2}^{f}m_{h}^{2}\right)+\left(e_{3}^{f}m_{h}^{3}\right)\,,\label{eq:I2}\\
I_{4}^{\left(1\right)} & = & -\left(-m_{f}^{0}e_{0}^{h}\right)\left(e_{1}^{f}m_{h}^{1}\right)+\left(-m_{f}^{0}e_{0}^{h}\right)\left(e_{2}^{f}m_{h}^{2}\right)+\left(e_{1}^{f}m_{h}^{1}\right)\left(e_{3}^{f}m_{h}^{3}\right)\nonumber \\
 &  & -\left(e_{2}^{f}m_{h}^{2}\right)\left(e_{3}^{f}m_{h}^{3}\right)\,,\label{eq:I4a}\\
I_{4}^{\left(2\right)} & = & -\left(-m_{f}^{0}e_{0}^{h}\right)\left(e_{2}^{f}m_{h}^{2}\right)+\left(-m_{f}^{0}e_{0}^{h}\right)\left(e_{3}^{f}m_{h}^{3}\right)+\left(e_{1}^{f}m_{h}^{1}\right)\left(e_{2}^{f}m_{h}^{2}\right)\nonumber \\
 &  & -\left(e_{1}^{f}m_{h}^{1}\right)\left(e_{3}^{f}m_{h}^{3}\right)\,,\label{eq:I4b}\end{eqnarray}
\begin{eqnarray}
I_{6} & = & \left(-m_{f}^{0}e_{0}^{h}\right)^{2}\left(e_{1}^{f}m_{h}^{1}\right)+\left(-m_{f}^{0}e_{0}^{h}\right)\left(e_{1}^{f}m_{h}^{1}\right)^{2}+\left(e_{2}^{f}m_{h}^{2}\right)^{2}\left(e_{3}^{f}m_{h}^{3}\right)+\left(e_{2}^{f}m_{h}^{2}\right)\left(e_{3}^{f}m_{h}^{3}\right)^{2}\nonumber \\
 &  & -4\left(e_{1}^{f}m_{h}^{1}\right)^{2}\left(e_{2}^{f}m_{h}^{2}\right)-4\left(e_{1}^{f}m_{h}^{1}\right)\left(e_{2}^{f}m_{h}^{2}\right)^{2}-4\left(-m_{f}^{0}e_{0}^{h}\right)^{2}\left(e_{3}^{f}m_{h}^{3}\right)-4\left(-m_{f}^{0}e_{0}^{h}\right)\left(e_{3}^{f}m_{h}^{3}\right)^{2}\nonumber \\
 &  & +3\left(-m_{f}^{0}e_{0}^{h}\right)\left(e_{1}^{f}m_{h}^{1}\right)\left(e_{2}^{f}m_{h}^{2}\right)+3\left(-m_{f}^{0}e_{0}^{h}\right)\left(e_{1}^{f}m_{h}^{1}\right)\left(e_{3}^{f}m_{h}^{3}\right)\nonumber \\
 &  & +3\left(-m_{f}^{0}e_{0}^{h}\right)\left(e_{2}^{f}m_{h}^{2}\right)\left(e_{3}^{f}m_{h}^{3}\right)+3\left(e_{1}^{f}m_{h}^{1}\right)\left(e_{2}^{f}m_{h}^{2}\right)\left(e_{3}^{f}m_{h}^{3}\right)\,.\label{eq:-I6}\end{eqnarray}
Note that given the sign restrictions in \eqref{eq:i-restrict}, each
term in parentheses is positive-definite. Also, note that exactly
four distinct \emph{products }of\emph{ }pairs of fluxes appear in
the expressions for the invariants. Duality orbits of our reduced
fluxes therefore sweep out a codimension 0 volume in the full space
of fluxes. It is more difficult to say whether pairs of fluxes satisfying
the sign constraints \eqref{eq:i-restrict} span the physically allowed
values of the invariants \eqref{eq:I2}-\eqref{eq:-I6}. 

The explicit form \eqref{eq:G-solution} of the generating function
$\mathcal{G}$ raises an interesting question. Three independent signs
appear, $\mbox{sgn}\left(m_{f}^{0}m_{h}^{1}\right),$ $\mbox{sgn}\left(m_{f}^{0}m_{h}^{2}\right),$
and $\mbox{sgn}\left(m_{f}^{0}m_{h}^{3}\right).$ One can readily
verify that duality transformations that leave the subspace of reduced
fluxes invariant also leave these signs, and only these signs, invariant.
Although we are not certain that these signs lift to invariants of
the full $SL\left(2\right)^{4},$ it is possible that they label different
octants of the full space of fluxes, with distinct expressions for
e.g. the gravitino mass in each octant.

We can use these facts to generalize our solution of the $F=Z^{1}Z^{2}Z^{3}/Z^{0}$
model with eight fluxes to a solution with all sixteen fluxes. We
propose the following procedure:
\begin{enumerate}
\item Consider \eqref{eq:I2}-\eqref{eq:-I6} to be a set of implicit functions
for each pair of fluxes in terms of $I_{2}=\int F_{3}\wedge H_{3},$
$I_{4}^{\left(1,2\right)},$ and $I_{6}.$
\item Substitute these functions into \eqref{eq:G-solution} to get $\mathcal{G}$
as a function of the invariants.
\item Substitute the full expressions for $I_{2}=\int F_{3}\wedge H_{3},$
$I_{4}^{\left(1,2\right)},$ and $I_{6}$ into $\mathcal{G}$ to get
an expression for $\mathcal{G}$ as a function of general fluxes.
\item Derivatives of $\mathcal{G}$ with respect to the fluxes will then
give the potentials, and in turn the values of the complex structure
moduli and mass parameters.
\item Solve \eqref{eq:3-0-Constraint} to determine the value of $\tau.$ 
\end{enumerate}
This procedure will certainly work if the eight additional fluxes
are small. As they become large, global properties of the space of
fluxes may present an obstruction, for example one of $\mbox{sgn}\left(m_{f}^{0}m_{h}^{1}\right),$
$\mbox{sgn}\left(m_{f}^{0}m_{h}^{2}\right),$ or $\mbox{sgn}\left(m_{f}^{0}m_{h}^{3}\right)$
might effectively flip. It is also possible that there are other branches
of solutions that we have not identified. 

Though considerations of duality-invariance have not yet led us to
a complete solution of the flux attractor equations with $F=Z^{1}Z^{2}Z^{3}/Z^{0},$
we hope that future work will make our understanding of flux compactifications
on this geometry as detailed as the modern understanding of the STU
black hole.

\section{\label{sec:Conclusion}Thermodynamics, Stability, and the Landscape}

One of the goals of this paper was to determine how much of the analysis
of flux compactifications could be done directly on the space of input
fluxes. We demonstrated that local properties of the compactification
are completely determined by a single generating function $\mathcal{G}$
defined on the space of fluxes. Although we have been conservative
in describing $\mathcal{G}$ as a {}``generating function,'' we
hope that future analysis will reveal that it is a proper thermodynamic
function, and that we can think of the fluxes themselves as the parameters
of an underlying thermodynamic system. At the same time, we might
worry that our success in constructing $\mathcal{G}$ hinged only
on the Kähler structure of the moduli space, and that no thermodynamic
interpretation exists. We now outline some of the principal challenges
surrounding a thermodynamic interpretation of flux attractors.
\begin{description}
\item [{Is~$\mathcal{G}$~a~Thermodynamic~Function?}] Equations \eqref{eq:phih-G-notau}
and \eqref{eq:phif-G-notau} look like equilibrium relations between
the fluxes and their thermodynamic conjugates. In addition to equilibrium
relations, thermodynamic functions also obey a set of stability conditions.
For a sensible thermodynamic interpretation, we would require that
stable and unstable thermodynamic equilibria correspond to stable
and unstable minima of the traditional spacetime potential \eqref{eq:NoScalePotential}.
Here we find an apparent mismatch between the two Hessians. While
the field-theoretic mass matrix has $2n+2$ eigenvalues, the matrix
of second derivatives of $\mathcal{G}$ has $4n+4$ eigenvalues. For
guidance we might study the analogous issue in the black hole attractor.
There, the Hessian of the effective potential has $2n$ eigenvalues,
while the second derivatives of the entropy lead to $2n+2$ eigenvalues.
\item [{What~Kind~of~Thermodynamic~Function~is~$\mathcal{G}$?}] In
thermodynamic problems, the energy and the entropy are treated rather
differently. In particular, energies are \emph{minimized} at stable
equilibria, while entropies are \emph{maximized.} In other ensembles
the energy is mapped to a free energy and the entropy to a generalized
Massieu function, but free energies are still minimized and Massieu
functions are still maximized. The interpretation of $\mathcal{G}$
hinges on whether it is minimized, in which case it might be interpreted
as the tension of a dual domain wall \citep{Behrndt:2001mx}, or maximized, in which case
it could be interpreted as an entropy. Determining this requires that
we fix the overall sign of $\mathcal{G}.$ Doing this might be as
simple as requiring that $\mathcal{G}$ be positive for stable configurations,
but it could be more subtle.
\item [{What~Does~This~Imply~for~the~Landscape?}] If we can establish
that $\mathcal{G}$ is an entropy, it becomes quite natural to propose
$e^{\mathcal{G}}$ as a classical measure on the string theory landscape.
Presumably such a measure would be related to the number of microscopic
realizations of a given set of fluxes. We can go on to ask if there
are any geometries for which this measure becomes strongly peaked,
or whether consistency conditions (such as the tadpole constraint)
require that $\mathcal{G}$ be $\mathcal{O}\left(1\right).$ 
\end{description}
Clearly many potential obstacles lie between the generating function
introduced in this paper and a \emph{predictive} measure on the landscape.
However the prospect of such a measure is quite exciting, and so worthy
of some attention.

\acknowledgments

We thank Sera Cremonini for collaboration in the initial stages of
this project. It is a pleasure to acknowledge helpful discussions
with Konstantin Bobkov, Alejandra Castro, Gianguido Dall'Agata, Josh
Davis, Sergio Ferrara, Jason Kumar, Jan Louis, Gary Shiu, Eva Silverstein,
Alessandro Tomasiello, and Brett Underwood. FL thanks CERN for hospitality
during part of this project. This work was supported by the DoE under
grant DE-FG02-95ER40899. 

\appendix

\section{\label{sec:Scalar-Mass-Matrix}Scalar Mass Matrix in No-Scale Compactifications}

In this appendix we present an explicit computation of the scalar
mass matrix for no-scale compactifications.\textbf{ }

We divide the scalar potential into two terms as follows:\begin{eqnarray}
V_{\mathrm{tot}} & = & V+V_{0}\\
 & = & e^{K}g^{\alpha\overline{\beta}}D_{\alpha}W\overline{D_{\beta}W}+e^{K}\left(g^{a\overline{b}}D_{a}W\overline{D_{b}W}-3\left|W\right|^{2}\right).\end{eqnarray}
The indices $\alpha,\beta,\gamma...$ run over the complex structure
moduli $i,j,k...$ and axio-dilaton $\tau,$ and $a,b,...$ run over
the Kähler moduli. Because the superpotential is independent of the
Kähler moduli, their F-terms are \eqref{eq:Kahler-F-Terms}  \begin{equation}
D_{a}W=W\partial_{a}K\,.\end{equation}
The inverse metric is such that \begin{equation}
g^{a\overline{b}}\partial_{a}K\overline{\partial}_{\overline{b}}K=3\,,\label{eq:NoScaleCondition}\end{equation}
so that $V_{0}=0.$ The remaining term $V$ is positive semi-definite,
so the absolute minima of the scalar potential all have vanishing
cosmological constant. This is why these solutions are called {}``no-scale.''

Since $V_{0}=0,$ we do not expect this term to make a contribution
to the mass matrix. We now show explicitly that this is the case,
beginning with the contribution to $M_{\alpha\beta}^{2}$ from $V_{0}:$\begin{eqnarray*}
\partial_{\beta}\partial_{\alpha}V_{0} & = & \partial_{\beta}\left\{ e^{K}\left[g^{a\overline{b}}\left(D_{\alpha}D_{a}W\overline{D_{b}W}+D_{a}WD_{\alpha}\overline{D_{b}W}\right)+D_{a}W\overline{D_{b}W}\partial_{\alpha}g^{a\overline{b}}\right]-3\overline{W}D_{\alpha}W\right\} \\
 & = & e^{K}\left[g^{a\overline{b}}\left(D_{\beta}D_{\alpha}D_{a}W\overline{D_{b}W}+D_{\alpha}D_{a}WD_{\beta}\overline{D_{b}W}\right)+\left(\partial_{\beta}g^{a\overline{b}}\right)D_{\alpha}D_{a}W\overline{D_{b}W}\right.\\
 &  & +g^{a\overline{b}}\left(D_{\beta}D_{a}WD_{\beta}\overline{D_{b}W}+D_{a}WD_{\beta}D_{\alpha}\overline{D_{b}W}\right)+\left(\partial_{\beta}g^{a\overline{b}}\right)D_{a}WD_{\alpha}\overline{D_{b}W}\\
 &  & +\left.\partial_{\alpha}g^{a\overline{b}}\left(D_{\beta}D_{a}W\overline{D_{b}W}+D_{a}WD_{\beta}\overline{D_{b}W}\right)+D_{a}W\overline{D_{b}W}\partial_{\beta}\partial_{\alpha}g^{a\overline{b}}-3\overline{W}D_{\beta}D_{\alpha}W\right].\end{eqnarray*}
Since the Kähler potential factorizes into $K=K_{z}\left(z^{i},\overline{z}^{\overline{i}}\right)+K_{\tau}\left(\tau,\overline{\tau}\right)+K_{t}\left(t^{a},\overline{t}^{\overline{a}}\right)$
we find that $\partial_{\alpha}g^{a\overline{b}}=0,$ and simplify further:
\begin{eqnarray}
\partial_{\beta}\partial_{\alpha}V_{0} & = & e^{K}\left[g^{a\overline{b}}\left(D_{\beta}D_{\alpha}D_{a}W\overline{D_{b}W}+D_{\alpha}D_{a}WD_{\beta}\overline{D_{b}W}\right)\right.\nonumber \\
 &  & +\left.g^{a\overline{b}}\left(D_{\beta}D_{a}WD_{\beta}\overline{D_{b}W}+D_{a}WD_{\beta}D_{\alpha}\overline{D_{b}W}\right)-3\overline{W}D_{\beta}D_{\alpha}W\right].\end{eqnarray}
Since $\partial_{\alpha}\partial_{a}K=0,$ we have \begin{eqnarray}
D_{\alpha}D_{a}W & = & D_{\alpha}\left(W\partial_{a}K\right)\\
 & = & \left(D_{\alpha}W\right)\partial_{a}K\end{eqnarray}
and\begin{eqnarray}
D_{\alpha}\overline{D_{b}W} & = & D_{\alpha}\left(\overline{W}\overline{\partial}_{b}K\right)\\
 & = & 0\,.\end{eqnarray}
This, combined with \eqref{eq:NoScaleCondition}, gives\begin{eqnarray}
\partial_{\beta}\partial_{\alpha}V_{0} & = & e^{K}\left[\left(g^{a\overline{b}}\partial_{a}K\overline{\partial}_{\overline{b}}K\right)\overline{W}D_{\beta}D_{\alpha}W-3\overline{W}D_{\beta}D_{\alpha}W\right]\\
 & = & 0\,,\end{eqnarray}
so $V_{0}$ indeed makes no contribution to $M_{\alpha\beta}^{2}.$ 

The contributions to $M_{\alpha\overline{\beta}}^{2}$ from $V_{0}$ simplify
in a similar way:\begin{eqnarray}
\overline{\partial}_{\overline{\beta}}\partial_{\alpha}V_{0} & = & \overline{\partial}_{\overline{\beta}}\left\{ e^{K}\left[g^{a\overline{b}}\left(D_{\alpha}D_{a}W\overline{D_{b}W}+D_{a}WD_{\alpha}\overline{D_{b}W}\right)+D_{a}W\overline{D_{b}W}\partial_{\alpha}g^{a\overline{b}}\right]-3\overline{W}D_{\alpha}W\right\} \nonumber \\
 & = & e^{K}\left[g^{a\overline{b}}\left(\overline{D}_{\overline{\beta}}D_{\alpha}D_{a}W\overline{D_{b}W}+D_{\alpha}D_{a}W\overline{D_{\beta}D_{b}W}\right)\right.\nonumber \\
 &  & \left.-3\left(\overline{W}\overline{D}_{\overline{\beta}}D_{\alpha}W+D_{\alpha}W\overline{D_{\beta}W}\right)\right]\\
 & = & 0\,.\end{eqnarray}
So our expectations were correct, and $V_{0}$ makes no contribution
to the scalar mass matrix. 

We emphasize that in computing the contributions from $V_{0}$ to
the mass matrix we have not set $D_{\alpha}W=0,$ we have only used
the factorization of the Kähler potential. Our conclusion that $V_{0}$
makes no contribution to the scalar mass matrix thus holds for metastable
local minima, where $D_{\alpha}W\neq0,$ as well as absolute minima,
where $D_{\alpha}W=0.$

Next we compute the contributions to the mass matrix from $V.$ Since
we are interested in absolute minima of the potential, we will set
$D_{\alpha}W=0.$ We begin with contributions to $M_{\alpha\beta}^{2}:$\begin{eqnarray}
\partial_{\beta}\partial_{\alpha}V & = & \partial_{\beta}\left\{ e^{K}\left[g^{\gamma\overline{\delta}}\left(D_{\alpha}D_{\gamma}W\overline{D_{\delta}W}+D_{\gamma}WD_{\alpha}\overline{D_{\delta}W}\right)+D_{\gamma}W\overline{D_{\delta}W}\partial_{\alpha}g^{\gamma\overline{\delta}}\right]\right\} \nonumber \\
 & = & e^{K}g^{\gamma\overline{\delta}}\left(D_{\alpha}D_{\gamma}WD_{\beta}\overline{D_{\delta}W}+D_{\beta}D_{\gamma}WD_{\alpha}\overline{D_{\delta}W}\right)\label{eq:Ugly1}\end{eqnarray}
We can eliminate the mixed derivatives using\begin{eqnarray}
D_{\alpha}\overline{D_{\beta}W} & = & D_{\alpha}\left(\overline{\partial_{\beta}W}+\overline{W}\overline{\partial}_{\overline{\beta}}K\right)\\
 & = & \overline{W}\partial_{\alpha}\overline{\partial}_{\overline{\beta}}K\\
 & = & \overline{W}g_{\alpha\overline{\beta}}\,,\end{eqnarray}
so that \eqref{eq:Ugly1} simplifies to \begin{eqnarray}
M_{\alpha\beta}^{2}=\partial_{\beta}\partial_{\alpha}V & = & e^{K}g^{\gamma\overline{\delta}}\left[D_{\alpha}D_{\gamma}WD_{\beta}\overline{D_{\delta}W}+D_{\beta}D_{\gamma}WD_{\alpha}\overline{D_{\delta}W}\right]\\
 & = & e^{K}\overline{W}\left(D_{\alpha}D_{\beta}W+D_{\beta}D_{\alpha}W\right).\label{eq:Mab}\end{eqnarray}
We'll follow the same procedure for $M_{\alpha\overline{\beta}}^{2},$
\begin{eqnarray}
M_{\alpha\overline{\beta}}^{2}=\overline{\partial}_{\overline{\beta}}\partial_{\alpha}V & = & \overline{\partial}_{\overline{\beta}}\left\{ e^{K}\left[g^{\gamma\overline{\delta}}\left(D_{\alpha}D_{\gamma}W\overline{D_{\delta}W}+D_{\gamma}WD_{\alpha}\overline{D_{\delta}W}\right)+D_{\gamma}W\overline{D_{\delta}W}\partial_{\alpha}g^{\gamma\overline{\delta}}\right]\right\} \nonumber \\
 & = & e^{K}g^{\gamma\overline{\delta}}\left[D_{\alpha}D_{\gamma}W\overline{D_{\beta}D_{\delta}W}+\overline{D}_{\overline{\beta}}D_{\gamma}WD_{\alpha}\overline{D_{\delta}W}\right]\nonumber \\
 & = & e^{K}\left[g^{\gamma\overline{\delta}}D_{\alpha}D_{\gamma}W\overline{D_{\beta}D_{\delta}W}+\left|W\right|^{2}g_{\alpha\overline{\beta}}\right].\label{eq:Mabbar}\end{eqnarray}
Our results for the scalar mass matrices, \eqref{eq:Mab} and \eqref{eq:Mabbar},
agree with the standard results for $\mathcal{N}=1$ supergravity,
e.g. eq. 23.27 in \citep{Wess:1992cp}. We have verified that the
Kähler moduli do not make any additional contributions.

We also see that when $W\neq0,$ i.e. when SUSY is broken, the scalar
masses-squared are lifted above the fermion masses-squared by $\mathcal{O}\left(m_{3/2}^{2}\right).$ 

\small
\bibliographystyle{hunsrt-abbrv}
\bibliography{RCO-FluxAttractors}

\end{document}